\documentclass[useAMS,usenatbib]{mn2e}

\usepackage{psfig}
\usepackage{graphicx}
\usepackage{times}
\usepackage{natbib}
\usepackage[latin1]{inputenc}

\newif\ifAMStwofonts
\AMStwofontstrue

\newcommand{\be}{\begin{equation}}
\newcommand{\ee}{\end{equation}}
\newcommand{\ba}{\begin{eqnarray}}
\newcommand{\ea}{\end{eqnarray}}
\newcommand{\brr}{\begin{array}}
\newcommand{\err}{\end{array}}
\newcommand{\bc}{\begin{center}}
\newcommand{\ec}{\end{center}}
\newcommand{\hm}{\,h^{-1}{\rm Mpc}}
\newcommand{\hk}{\,h^{-1}{\rm kpc}}
\newcommand{\msun}{\,h^{-1}M_\odot}

\newcommand{\vel}{\,{\rm km\,s^{-1}}}

\newcommand{\Zfe}{\mbox{$Z_{\rmn{Fe}} \,$}}

\newcommand{\Zsife}{\mbox{$Z_{\rmn{Si}}/Z_{\rmn{Fe}}~$}}

\newcommand{\gadget}{{\footnotesize {\sc GADGET-2~}}}

\pssilent 

\newcommand{\mincir}{\raise
  -2.truept\hbox{\rlap{\hbox{$\sim$}}\raise5.truept \hbox{$<$}\ }}
\newcommand{\magcir}{\raise
  -2.truept\hbox{\rlap{\hbox{$\sim$}}\raise5.truept \hbox{$>$}\ }}
\newcommand{\siml}{\raise
  -2.truept\hbox{\rlap{\hbox{$\sim$}}\raise5.truept \hbox{$<$}\ }}
\newcommand{\simg}{\raise
  -2.truept\hbox{\rlap{\hbox{$\sim$}}\raise5.truept \hbox{$>$}\ }}

\title[AGN feedback and metal enrichment of clusters] {Simulating the
  effect of AGN feedback on the metal enrichment of galaxy clusters}
\author[Fabjan et al.] {D. Fabjan$^{1,2,3}$, S. Borgani$^{1,2,3}$,
  L. Tornatore$^{1,2,3}$, A. Saro$^{1,2,3}$, G. Murante$^4$ \& K. Dolag$^5$\\
  ~\\
  $^1$ Dipartimento di Astronomia dell'Universit\`a di Trieste, via
  Tiepolo 11, I-34131 Trieste, Italy (fabjan,borgani,tornatore,saro@oats.inaf.it)\\
  $^2$ INAF -- Istituto Nazionale di Astrofisica, via Tiepolo 11,
  I-34131 Trieste, Italy\\
  $^3$ INFN -- Istituto Nazionale di Fisica Nucleare, Trieste, Italy\\
  $^4$ INAF -- Istituto Nazionale di Astrofisica -- Osservatorio
  Astronomico di Torino, Str. Osservatorio 25, I-10025, Pino Torinese,
  Torino, Italy (murante@oato.inaf.it)\\
  $^5$ Max-Planck-Institut f\"ur Astrophysik, Karl-Schwarzschild
  Strasse
  1, Garching bei M\"unchen, Germany (kdolag@mpa-garching.mpg.de)\\
}

\begin{document}
\label{firstpage}
\maketitle

\begin{abstract}
  We present a study of the effect of AGN feedback on metal enrichment
  and thermal properties of the intracluster medium (ICM) in
  hydrodynamical simulations of galaxy clusters. The simulations are
  performed using a version of the TreePM-SPH \gadget code, which also
  follows chemo-dynamical evolution by accounting for metal enrichment
  contributed by different stellar populations. We carry out
  cosmological simulations for a set of galaxy clusters, covering the
  mass range $M_{200}\simeq (0.1-2.2)\times 10^{15}\msun$. Besides
  runs not including any efficient form of energy feedback, we carry
  out simulations including three different feedback schemes: {\em
    (i)} kinetic feedback in the form of galactic winds triggered by
  supernova explosions; {\em (ii)} AGN feedback from gas accretion
  onto super-massive black holes (BHs); {\em (iii)} AGN feedback in
  which a 'radio mode' is included with an efficient thermal coupling
  of the extracted energy, whenever BHs enter in a quiescent accretion
  phase. Besides investigating the resulting thermal properties of the
  ICM, we analyse in detail the effect that these feedback models have
  on the ICM metal-enrichment. We find that AGN feedback has the
  desired effect of quenching star formation in the brightest cluster
  galaxies at $z<4$ and provides correct temperature profiles in the
  central regions of galaxy groups. However, its effect is not yet
  sufficient to create ``cool cores'' in massive clusters, while
  generating an excess of entropy in central regions of galaxy
  groups. As for the pattern of metal distribution, AGN feedback
  creates a widespread enrichment in the outskirts of clusters, thanks
  to its efficiency in displacing enriched gas from galactic halos to
  the inter-galactic medium. This turns into profiles of Iron
  abundance, \Zfe, which are in better agreement with observational
  results, and into a more pristine enrichment of the ICM around and
  beyond the cluster virial regions. Following the pattern of the
  relative abundances of Silicon and Iron, we conclude that a
  significant fraction of the ICM enrichment is contributed in
  simulations by a diffuse population of intra-cluster stars. Our
  simulations also predict that profiles of the \Zsife abundance ratio
  do not increase at increasing radii, at least out to
  $0.5R_{vir}$. Our results clearly show that different sources of
  energy feedback leave distinct imprints in the enrichment pattern of
  the ICM. They further demonstrate that such imprints are more
  evident when looking at external regions, approaching the cluster
  virial boundaries.
\end{abstract}


\section{Introduction}
High quality data from the current generation of X--ray satellites
(XMM-Newton, Chandra and Suzaku) have now established a number of
observational facts concerning the thermo-dynamical and
chemo-dynamical properties of the intra-cluster medium (ICM) for
statistically representative sets of galaxy clusters: core regions of
relaxed clusters show little evidence of gas cooler than about a third
of the virial temperature
\citep[e.g.,][]{Peterson2001A&A...365L.104P,Boehringer2002A&A...382..804B,
Sanderson2006MNRAS.372.1496S}; temperature profiles have negative
gradients outside core regions, a trend that extends out to the largest
radii covered so far by observations
\citep[e.g., ][]{DeGrandi2002ApJ...567..163D,Vikhlinin2005ApJ...628..655V,Zhang2006A&A...456...55Z,
Baldi2007ApJ...666..835B,Pratt2007A&A...461...71P,Leccardi2008A&A...486..359L};
gas entropy is higher than expected from simple self-similar scaling
properties of the ICM, not only in core regions, but also out to
$R_{500}$\footnote{Here and in the following we will indicate with
  $R_\Delta$ the radius encompassing a mean density equal to $\Delta
  \rho_{cr}$, being $\rho_{cr}=3H^2/8\pi G$ the critical cosmic
  density. In a similar way, we will indicate with $M_\Delta$ the mass
  contained within $R_\Delta$.}  \citep[e.g.,][and references
therein]{Sun2009ApJ...693.1142S}; radial profiles of the Iron
abundance show negative gradients, more pronounced for relaxed
``cool core'' clusters, with central values of \Zfe approaching solar
abundance and with a global enrichment at a level of about 1/3--1/2
$Z_{Fe,\odot}$
(e.g.,\citealt{DeGrandi2004A&A...419....7D,Vikhlinin2005ApJ...628..655V,
  dePlaa2006A&A...452..397D,Snowden2008A&A...478..615S,Leccardi2008A&A...487..461L};
see \citealt{Mushotzky2004cgpc.symp..123M,Werner2008SSRv..134..337W}
for recent reviews).

Galaxy clusters arise from the collapse of density perturbations over
a scale of $\sim 10\,h^{-1}$ comoving Mpc. As such, the above observational
properties of the ICM come from a non-trivial interplay between the
underlying cosmological scenario, which shapes the large-scale
structure of the Universe, and a number of astrophysical processes
(e.g., star formation, energy and chemical feedback from supernovae
and AGN) taking place on much smaller scales. Cosmological
hydrodynamical simulations represent the modern instrument with which
such complexities can be described as the result of hierarchical
assembly of cosmic structures \citep[e.g.,][for a recent
review]{Borgani2009arXiv0906.4370B}. Indeed, simulations of galaxy
clusters reach nowadays a high enough resolution, while including a
realistic description of the above mentioned astrophysical processes,
for them to provide a coherent interpretative framework for X-ray
observations. Quite remarkably, simulation predictions for the
thermodynamical properties of the ICM are in good agreement with
observations, at least outside the core regions: simulated profiles of
gas density and temperature match the observed ones at cluster-centric
distances $\magcir 0.1R_{500}$
\citep[e.g.,][]{Loken2002ApJ...579..571L,Borgani2004MNRAS.348.1078B,
  Kay2004MNRAS.355.1091K,Roncarelli2006MNRAS.373.1339R,
  Pratt2007A&A...461...71P,Nagai2007ApJ...655...98N,Croston2008A&A...487..431C};
the observed entropy level at $R_{500}$ is well reproduced by
simulations including radiative cooling and star formation
\citep[e.g.,][]{Nagai2007ApJ...668....1N,Borgani2009MNRAS.392L..26B}.
The situation is quite different within cluster cores, where
simulations including only stellar feedback generally fail at
producing realistic cool cores. The two clearest manifestations of
this failure are represented by the behaviour of temperature and
entropy profiles at small radii, $\mincir 0.1R_{500}$. Observations of
cool core clusters show that temperature profiles smoothly decline
toward the centre, reaching temperatures of about 1/3--1/2 of the
maximum value, while the entropy level at the smallest sampled scales
is generally very low
\citep[e.g.,][]{Sun2009ApJ...693.1142S,Sanderson2009MNRAS.tmp..428S}. On
the contrary, radiative simulations including a variety of models of
stellar feedback predict steep negative temperature gradients down to
the innermost resolved radii and central entropy levels much higher
than observed (e.g.,
\citealt{Valdarnini2003MNRAS.339.1117V,Tornatore2003MNRAS.342.1025T,
  Borgani2004MNRAS.348.1078B,Nagai2007ApJ...668....1N}; cf. also
\citealt{Kay2007MNRAS.377..317K}). This failure of simulations is
generally interpreted as due to overcooling, which takes place in
simulated clusters even when including an efficient supernova (SN)
feedback, and causes an excess of star formation in the
simulated brightest cluster galaxies (BCGs;
\citealt{Romeo2005MNRAS.361..983R,Saro2006MNRAS.373..397S}).

The generally accepted solution to these shortcomings of simulations
is represented by AGN feedback. Indeed, the presence of cavities in
the ICM at the cluster centre is considered as the fingerprint of the
conversion of mechanical energy associated to AGN jets into
thermal energy (and possibly in a non--thermal content of relativistic
particles) through shocks (e.g.,
\citealt{Mazzotta2004JKAS...37..381M,Fabian2005MNRAS.360L..20F,
  McNamara2006ApJ...648..164M, Sanders2007MNRAS.381.1381S}; see
\citealt{McNamara2007ARA&A..45..117M} for a review). Although
analytical arguments convincingly show that the energy radiated from
gas accretion onto a central super-massive black hole (BH) is enough to
suppress gas cooling, it is all but clear how this energy is
thermalised and distributed in the surrounding medium. A likely
scenario is that bubbles of high entropy gas are created at the
termination of jets. Buoyancy of these bubbles in the ICM then
distributes thermal energy over larger scales
\citep[e.g.,][]{DallaVecchia2004MNRAS.355..995D,Cattaneo2007MNRAS.377...63C}.
Crucial in this process is the stability of the bubbles against gas
dynamical instabilities which would tend to destroy them quite
rapidly. Indeed, detailed numerical simulations have demonstrated that
gas circulation associated to jets
\cite[e.g.,][]{Omma2004MNRAS.348.1105O,Brighenti2006ApJ...643..120B,Heinz2006MNRAS.373L..65H},
gas viscosity, magnetic fields
\citep[e.g.,][]{Ruszkowski2007MNRAS.378..662R} and injection of cosmic
rays \citep[e.g.,][]{Ruszkowski2008MNRAS.383.1359R} are all expected
to cooperate in determining the evolution of buoyant bubbles. Although
highly instructive, all these simulations have been performed for
isolated halos and, as such, they describe neither the cosmological
growth and merging of black holes, nor the hierarchical assembly of
galaxy clusters.

\cite{Springel2005MNRAS.361..776S} and
\cite{DiMatteo2005Natur.433..604D} have presented a model that follows
in a cosmological simulation the evolution of the BH population and
the effect of energy feedback resulting from gas accretion onto
BHs. In this model, BHs are treated as sink particles which accrete
from the surrounding gas according to a Bondi accretion rate
\citep[e.g.,][]{Bondi1952MNRAS.112..195B}, with an upper limit
provided by the Eddington rate \citep[see
also][]{Booth2009arXiv0904.2572B}. This model was used by
\cite{Bhattacharya2008MNRAS.389...34B}, who run cosmological
simulations to study the effect of AGN feedback on galaxy groups. They
found that this feedback is effective in reducing star formation in
central regions and to displace gas towards outer regions. However,
the resulting entropy level within cluster cores is higher than
observed. In its original version, the energy extracted from BH
accretion is locally distributed to the gas around the BHs according
to a SPH-kernel weighting scheme. This model has been modified by
\cite{Sijacki2006MNRAS.366..397S} who included the possibility to
inflate high-entropy bubbles in the ICM whenever accretion onto the
central BH enters in a quiescent ``radio mode''. The underlying idea
of injecting bubbles with this prescription was to provide a more
realistic description of the effect of jet termination on the ICM,
although the effect of the jet itself was not
included. \cite{Puchwein2008ApJ...687L..53P} simulated a set of
clusters and groups to show that this feedback scheme is able to
reproduce the observed slope of the relation between X--ray luminosity
and temperature. \cite{Sijacki2007MNRAS.380..877S} showed in a
simulation of a single poor galaxy cluster that the injection of
bubbles is quite effective in suppressing star formation in central
regions. However, also in this case the temperature profile in the
core regions does not match the observed slope, while more realistic
temperature profiles can be produced if bubble injection is associated
to the injection of a non--thermal population of relativistic
particles
\citep{Pfrommer2007MNRAS.378..385P,Sijacki2008MNRAS.387.1403S}.
Although these authors presented results concerning the effect of AGN
on the ICM thermodynamics, no detailed analysis has been so far
carried out to study the interplay between AGN and chemical
enrichment, by including a detailed chemo-dynamical description of the
ICM.

Indeed, the amount and distribution of metals in the ICM
provide an important diagnostic to reconstruct the past history of
star formation and the role of gas-dynamical and feedback processes in
displacing metals from star forming regions \citep[e.g.,][for
reviews]{Mushotzky2004cgpc.symp..123M,Borgani2008SSRv..134..379B,
  Schindler2008SSRv..134..363S}. Cosmological chemo-dynamical
simulations of galaxy clusters generally show that the predicted 
profiles of Iron abundance are steeper than observed
\citep[e.g.,][]{Valdarnini2003MNRAS.339.1117V,Tornatore2004MNRAS.349L..19T,
  Romeo2005MNRAS.361..983R,Tornatore2007MNRAS.382.1050T,Dave2008MNRAS.391..110D},
with an excess of enrichment in the core regions. This is generally
interpreted as due to the same excess of recent star formation in
simulated BCGs. However, \cite{Fabjan2008MNRAS.386.1265F} showed that
an excess of recent star formation has also the effect of efficiently locking
recently-produced metals into stars, thereby preventing
a too fast increase of the metallicity of the hot diffuse medium.

All the above analyses are based on different implementations of SN
energy feedback, while only much less detailed analyses of ICM metal
enrichment have been so far presented by also including AGN feedback
in cosmological simulations
\citep[e.g.,][]{Sijacki2007MNRAS.380..877S,Moll2007A&A...463..513M}. For
instance, \cite{Roediger2007MNRAS.375...15R} showed from simulations
of isolated halos that buoyancy of bubbles can actually displace a
large amount of the central highly enriched ICM, thus leading to a
radical change of the metallicity profiles, or even to a disruption of
the metallicity gradients.

The aim of this paper is to present a detailed analysis of
cosmological hydrodynamical simulations of galaxy clusters, which have
been carried out with the \gadget code
\citep{Springel2005MNRAS.364.1105S}, by combining the AGN feedback
model described by \cite{Springel2005MNRAS.361..776S} with the SPH
implementation of chemo-dynamics presented by
\cite{Tornatore2007MNRAS.382.1050T}. Besides showing results on the
effect of combining metallicity--dependent cooling and AGN feedback on
the ICM thermodynamics, we will focus our discussion on the different
effects that SNe and AGN feedback have on the chemical enrichment of the
ICM. Although we will shortly discuss the effect of different feedback
sources on the pattern of star formation in the BCG, the results
presented in this paper will mainly concern the enrichment of the hot
diffuse intra-cluster gas. We defer to a forthcoming paper a detailed
analysis of the effect of AGN feedback on the population of cluster
galaxies. The scheme of the paper is as follows. We present in Section
2 the simulated clusters, and briefly describe the implementation of
the chemical evolution model in the \gadget code along with the
SN and AGN feedback models. In Section 3 we show our results on
the thermal properties of the ICM and their comparison with
observational results. Section 4 is devoted to the presentation of the
results on the ICM chemical enrichment from our simulations and their
comparison with the most recent observations. We discuss our results
and draw our main conclusions in Section 5.

\section{The simulations}
\subsection{The set of simulated clusters}
Our set of simulated clusters is obtained from nine Lagrangian
regions extracted from a simulation containing only dark matter (DM)
with a box size of 479 h$^{-1}$ Mpc
\citep{Yoshida2001MNRAS.328..669Y}, performed for a flat $\Lambda$
cold dark matter (CDM) cosmological model with relevant parameters
$\Omega_m = 0.3$, h$_{100}=0.7$, $\sigma_8=0.9$. This set of simulated
clusters is described in detail by
\cite{Dolag2008arXiv0808.3401D}. The regions are centred on nine
galaxy clusters: five of them have masses $M_{200}\simeq 1.0 \times
10^{14} \msun$, while the remaining four have $M_{200}\simeq (1.0-2.2)
\times 10^{15} \msun$. The regions of three largest objects also
contain other clusters, so that we have in total 18 clusters with
$M_{200}\magcir 5\times 10^{13}\msun$, whose basic characteristics are
reported in Table \ref{t:clus}.

Each region was re-simulated using the zoomed initial condition (ZIC)
technique by \cite{Tormen1997MNRAS.286..865T}. The relative mass of
the gas and dark matter (DM) particles within the high--resolution
region of each simulation is set so that $\Omega_{bar}=0.045$ for the
density parameter contributed by baryons. For each Lagrangian region,
initial conditions have been generated at a basic resolution and at
$6.5$ times higher mass resolution. At this higher resolution, the
mass of the DM and gas particles are $m_{DM} \simeq 1.9 \times 10^8
\msun$ and $m_{gas} = 2.8 \times 10^7 \msun$ respectively, with
Plummer equivalent softening length for the computation of the
gravitational force set to $\varepsilon=2.75 \hk$ in physical units
below $z = 2$ and to $\varepsilon = 8.25 \hk$ in comoving units at
higher redshifts. In the lower resolution runs, these values are
rescaled according to $m_{DM}^{-1/3}$. We decided to simulate the nine
Lagrangian regions with massive clusters at the lower resolution,
while we used the higher resolution for the five low-mass clusters.
Only the low-mass g676 cluster was simulated at both resolutions. We
provide in Table \ref{t:res} a description of the parameters for the
two different resolutions, also listing the clusters simulated at each
resolution.

\begin{table} 
\centering
\caption{Characteristics of the simulated clusters. Column 1: cluster name.
  Column 2: mass contained within $R_{200}$ (units of
  $10^{14}\msun$). Column 3: value of $R_{200}$ (units of $\hm$). Column
  4: value of the spectroscopic-like temperature within $R_{500}$,
  $T_{500}$. Columns 5
  and 6: mass of the central BH hosted in the BCG for the AGN1 and
  AGN2 runs (units of $10^{10}\msun$), respectively. For the g51
    cluster, the two additional values reported for the mass of the central BH
    refer to the AGN2(0.8) and AGN2W runs (see text) and are indicated with the
    $\dagger$ and $\ddagger$ symbol, respectively.} 
\begin{tabular}{lrrrrr}
Cluster & $M_{200}$ & $R_{200}$ & $T_{500}$ & \multicolumn{2}{c}{$M_{BH}$}\\ 
   &  &  &  & AGN1 & AGN2 \\
\hline
g1.a  & 12.69 	& 1.76 	& 9.71 	& 56.22 	& 15.74		\\
g1.b  & 3.64  	& 1.16 	& 3.39 	& 7.96 		& 2.52 		\\
g1.c  & 1.36  	& 0.84 	& 2.03 	& 4.63 		& 1.32	 	\\
g1.d  & 1.04  	& 0.76 	& 1.82 	& 1.59 		& 0.46 		\\
g1.e  & 0.62  	& 0.64 	& 1.44 	& 0.84 		& 0.30		\\ ~\\
g8.a  & 18.51 	& 2.00 	& 12.82         & 113.50       	& 27.73 	\\
g8.b  & 0.65 	& 0.65 	& 1.41 	& 2.58 		& 0.88 		\\
g8.c  & 0.52 	& 0.61 	& 1.42 	& 1.74 		& 0.63		\\ ~\\
g51   & 10.95 	& 1.68 	& 7.55 	& 43.38 	        & 16.50		\\ 
      &         &               &               &
      &6.18$^\dagger$   \\
      &         &               &               &
      &15.02$^\ddagger$   \\
g72.a & 10.57 	& 1.66 	& 8.67 	& 48.74 	        & 13.87		\\
g72.b & 1.48 	& 0.86 	& 2.18 	& 3.79 		& 1.34		\\ ~\\
g676  & 0.87 	& 0.72 	& 1.91 	& 3.03 		& 0.95		\\
g914  & 0.88 	& 0.72 	& 2.03 	& 2.89 		& 1.13		\\
g1542 & 0.83 	& 0.71 	& 1.79 	& 2.54 		& 0.79		\\
g3344 & 0.92 	& 0.74 	& 2.03 	& 3.15 		& 1.29		\\
g6212 & 0.89 	& 0.73 	& 2.02 	& 2.57 		& 0.99		\\
\end{tabular}
\label{t:clus}
\end{table}

\begin{table} 
\centering
\caption{Resolution of the different runs.
  Column 2-3: mass of the DM particles and initial mass of the gas
  particles (units of 
  $10^{8}\msun$). Column 4: value of the Plummer-equivalent
  gravitational force softening at $z=0$ (units of $\hk$). Column
  5: clusters simulated at each resolution.} 
\begin{tabular}{lcccl}
 & $M_{DM}$ & $m_{gas}$ & $\varepsilon$ & Clusters \\ 
\hline 
Low res. & 11.0 & 2.03  & 5.00 & g1, g8, g51, g72, g676 \\
High res.& 1.69 & 0.31  & 2.75 & g676, g914, g1542, g3344, \\
         &      &       &      & g6212
\end{tabular}
\label{t:res}
\end{table}

\subsection{The simulation code}
Our simulations were performed using the TreePM-SPH \gadget code
\citep{Springel2005MNRAS.364.1105S}. All simulations include a
metallicity-dependent radiative cooling
\citep{Sutherland1993ApJS...88..253S}, heating from a uniform
time-dependent ultraviolet background
\citep{Haardt1996ApJ...461...20H} and the effective model by
\cite{Springel2003MNRAS.339..289S} for the description of star
formation. In this model, gas particles above a given density are
treated as multiphase, so as to provide a sub--resolution description
of the inter-stellar medium. In the following, we assume the density
threshold for the onset of star formation in multiphase gas particles
to be $n_H = 0.1 cm^{-3}$ in terms of number density of hydrogen
atoms. Our simulations also include a detailed model of chemical
evolution by \citeauthor{Tornatore2007MNRAS.382.1050T} (2007, T07
hereafter). We address the reader to T07 for a more detailed
description of this model. Metals are produced by SNe-II, SNe-Ia and
intermediate and low-mass stars in the asymptotic giant branch (AGB
hereafter). We assume SNe-II to arise from stars having mass above
$8M_\odot$. As for the SNe-Ia, we assume their progenitors to be binary
systems, whose total mass lies in the range (3--16)$M_\odot$. Metals
and energy are released by stars of different mass by properly
accounting for mass--dependent lifetimes. In this work we assume the
lifetime function proposed by \cite{Padovani1993ApJ...416...26Pb},
while we assume the standard stellar initial mass function (IMF) by
\cite{Salpeter1955ApJ...121..161S}. We adopt the
metallicity--dependent stellar yields by
\cite{Woosley1995ApJS..101..181W} for SNe-II, the yields by
\cite{vandenHoek1997A&AS..123..305V} for the AGB and by
\cite{Thielemann2003NuPhA.718..139T} for SNe-Ia. The version of the
code used for the simulations presented here allowed us to follow H,
He, C, O, Mg, S, Si and Fe. Once produced by a star particle, metals
are then spread to the surrounding gas particles by using the B-spline
kernel with weights computed over 64 neighbours and taken to be
proportional to the volume of each particle (see T07 for detailed
tests on the effect of choosing different weighting schemes to spread
metals).

\subsection{Feedback models}
In the simulations presented in this paper we model two different
sources of energy feedback. The first one is the kinetic feedback
model implemented by \cite{Springel2003MNRAS.339..289S}, in which
energy released by SN-II triggers galactic winds, with mass upload rate
assumed to be proportional to the star formation rate, $\dot{M}_W
=\eta \dot{M}_{\star}$. Therefore, fixing the parameter $\eta$ and the
wind velocity $v_W$ amounts to fix the total energy carried by the
winds. In the following, we assume $\eta=2$ for the mass-upload
parameter and $v_W = 500\vel$ for the wind velocity. If each SN-II
releases $10^{51}$ ergs, the above choice of parameters corresponds to
assuming that SNe-II power galactic outflows with nearly unity
efficiency for a Salpeter IMF \citep[see][]{Springel2003MNRAS.339..289S}.

Furthermore, we include in our simulations the effect of feedback
energy released by gas accretion onto super-massive black holes (BHs),
following the scheme originally introduced by
\citeauthor{Springel2005MNRAS.361..776S} (2005, SDH05 hereafter; see
also
\citealt{DiMatteo2005Natur.433..604D,DiMatteo2008ApJ...676...33D,Booth2009arXiv0904.2572B}).
We refer to SDH05 for a more detailed description. In this model, BHs
are represented by collisionless sink particles of initially very
small mass, that are allowed to subsequently grow via gas accretion
and through mergers with other BHs during close encounters.  During
the growth of structures, we seed every new dark matter halo above a
certain mass threshold $M_{th}$, identified by a run-time
friends-of-friends algorithm, with a central BH of mass $10^5 \msun$,
provided the halo does not contain any BH yet. For the runs at the
lower resolution the value of the halo mass assumed to seed BHs is
$M_{th}=5\times 10^{10}\msun$, so that it is resolved with about 40 DM
particles. At the higher resolution the halo mass threshold for BH seeding
decreases to $M_{th}=10^{10}\msun$, so as to resolve it with
approximately the same number of particles. We verified that using in
the higher-resolution runs the same value of $M_{th}$ as in the
low-resolution runs causes the effect of BH accretion to be shifted
toward lower redshift, since its onset has to await the formation of
more massive halos, while leaving the final properties of galaxy
clusters almost unaffected.

Once seeded, each BH can then grow by local gas accretion, with a rate
given by
\be
\dot M_{BH}\,=\,{\rm min}\left(\dot M_B, \dot M_{Edd}\right)\,
\label{eq:acrate}
\ee 
or by merging with other BHs.
Here $\dot M_B$ is the accretion rate estimated with the
Bondi-Hoyle-Lyttleton formula \citep{Hoyle1939PCPS...35..405H,
  Bondi1944MNRAS.104..273B,Bondi1952MNRAS.112..195B}, while $\dot
M_{Edd}$ is the Eddington rate. The latter is inversely proportional
to the radiative efficiency $\epsilon_r$, which gives the radiated
energy in units of the energy associated to the accreted mass:
\be
\epsilon_r\,=\,{L_r\over \dot M_{BH} c^2}\,.
\label{eq:radeff}
\ee 
Following \cite{Springel2005MNRAS.361..776S}, we use $\epsilon_r=0.1$
as a reference value, which is typical for a radiatively efficient
accretion onto a Schwartzschild BH \citep{Shakura1973A&A....24..337S}.
The model then assumes that a fraction $\epsilon_f$ of the radiated
energy is thermally coupled to the surrounding gas, so that $\dot
E_{\rm feed}=\epsilon_r \epsilon_f \dot M_{BH}c^2$ is the rate of
provided energy feedback. In standard AGN feedback
  implementation we use $\epsilon_f=0.05$ following
  \cite{DiMatteo2005Natur.433..604D}, who were able with this value
  to reproduce the observed $M_{BH}-\sigma$ relation between bulge
  velocity dispersion and mass of the hosted BH
  \citep[e.g.,][]{Magorrian1998AJ....115.2285M}. This choice was also
  found to be consistent with the value required in semi-analytical
  models to explain the evolution of the number density of quasars
  \citep{Wyithe2003ApJ...595..614W}.

Gas swallowed by the BH is implemented in a stochastic way, by
assigning to each neighbour gas particle a probability of contributing
to the accretion, which is proportional to the SPH kernel weight
computed at the particle position. Differently from SDH05, we assume
that each selected gas particle contributes to the accretion with 1/3
of its mass, instead of being completely swallowed. In this way, a
larger number of particles contribute to the accretion, which is then
followed in a more continuous way. We remind that in the SDH05
scheme, this stochastic accretion is used only to increase the dynamic
mass of the BHs, while their mass entering in the computation of the
accretion rate is followed in a continuous way, by using the analytic
expression for $\dot M_{BH}$. Once the amount of energy to be
thermalised is computed for each BH at a given time-step, one has to
distribute this energy to the surrounding gas particles. In their
original formulation, SDH05 distributed this energy using the SPH
kernel.

Besides following this standard implementation of the AGN feedback, we
also follow an alternative prescription, which differ from the
original one in two aspects.

Firstly, following \cite{Sijacki2007MNRAS.380..877S}, we assume that a
transition from a ``quasar'' phase to ``radio'' mode of the BH
feedback takes place whenever the accretion rate falls below a given
limit \citep[e.g.,][and references
therein]{Churazov2005MNRAS.363L..91C}, corresponding to $\dot
M_{BH}/\dot M_{Edd}< 10^{-2}$, which implies an increase of the
feedback efficiency to $\epsilon_f=0.2$. At high redshift BHs are
characterised by high accretion rates and power very luminous quasars,
with only a small fraction of the radiated energy being thermally
coupled to the surrounding gas. On the other hand, BHs hosted within
very massive halos at lower redshift are expected to accrete at a rate
well below the Eddington limit, while the energy is mostly released in
a kinetic form, eventually thermalised in the surrounding gas through
shocks. Secondly, instead of distributing the energy using a SPH
kernel, we distribute it using a top-hat kernel, having radius given
by the SPH smoothing length. In order to avoid spreading the energy
within a too small sphere, we assume a minimum spreading length of
$2\,h^{-1}$kpc. The rationale behind the choice of the top-hat kernel
is to provide a more uniform distribution of energy, thus
mimicking the effect of inflating bubbles in correspondence of the
termination of the AGN jets. It is well known that a number of
physical processes need to be adequately included for a fully
self-consistent description of bubble injection and buoyancy: 
gas-dynamical effects related to jets, magnetic fields, viscosity,
thermal conduction, injection of relativistic particles.

  In particular, a number of studies based on simulations of
  isolated halos
  \citep[e.g.,][]{Omma2004MNRAS.348.1105O,Brighenti2006ApJ...643..120B}
  have pointed out that gas circulation generated by jets provides an
  important contribution for the stabilization of cooling flows
  \citep[see also][for a cosmological simulation of cluster formation
  including jets]{Heinz2006MNRAS.373L..65H}. In its current
  implementation, the model of BH feedback included in our simulations
  neglects any kinetic feedback associated to jets. Based on an
  analytical model, \cite{Pope2009MNRAS.395.2317P} computed the
  typical scale of transition from kinetic to thermal feedback regime
  for AGN in elliptical galaxies and clusters. As a result, he found
  that the effect of momentum carried by jets can be neglected on
  scales $\magcir 20$ kpc, the exact value depending on the local
  conditions of the gas and on the injection rate of kinetic and
  thermal energy. In order to compare such a scale to that actually
  resolved in our simulations, we remind the reader that SPH hydrodynamics is
  numerically converged on scales about 6 times larger than the
  Plummer--equivalent softening scale for gravitational force
  \citep[e.g.,][]{Borgani2002MNRAS.336..409B}. Owing to the values of
  the gravitational softening reported in Table 2, the scales resolved
  in our simulations are not in the regime where kinetic feedback is
  expected to dominate, thus justifying the adoption of a purely
  thermal feedback. As a further test, we have computed the radius of
  the top--hat kernel within which energy is distributed around the
  central BHs in our simulated clusters. As a few examples, we found
  at $z=0$ this radius to be 21 kpc and 23 kpc for the AGN2 runs of
  the g72 and g676 clusters, respectively. This implies that we are in
  fact distributing thermal energy over scales where kinetic feedback
  should not be dominant.

In view of the difficulty of self--consistently including the
cooperative effect of all the physical processes we listed above, we
prefer here to follow a rather simplified approach and see to what
extent [results on] the final results of our simulations are sensitive
to variations in the implementation of the BH feedback model.

In summary, we performed four series of runs, corresponding to as many
prescription for energy feedback.
\begin{description}
\item[(a)] No feedback (NF hereafter): neither galactic winds nor AGN
  feedback is included.
\item[(b)] Galactic winds (W hereafter) included by following the
  model by \cite{Springel2003MNRAS.339..289S}, with $v_w=500\vel$ and
  $\eta=2$ for the wind mass upload.
\item[(c)] Standard implementation of AGN feedback from BH accretion
  (AGN1 hereafter), using $\epsilon_f=0.05$ for the feedback
  efficiency.
\item[(d)] Modified version of the AGN feedback (AGN2 hereafter),
  with feedback efficiency increasing from $\epsilon_f=0.05$ to
  $\epsilon_f=0.2$ when $\dot M_{BH}/\dot M_{Edd}< 10^{-2}$, and
  distribution of energy around the BH with a top-hat kernel.
\end{description}

  In order to further explore the parameter space of the considered
  feedback models we also carried out one simulation of the g51
  cluster based on the AGN2 scheme, but with the feedback efficiency
  increased to $\epsilon_f=0.8$ (AGN2(0.8) hereafter). We also note
  that our simulations include either winds triggered by SN explosions
  or AGN feedback. While this choice is done with the purpose of
  separate the effects of these two feedback sources, we expect in
  realistic cases that both AGN and SN feedback should cooperate in
  determining the star formation history of galaxies. In order to
  verify the effect of combining the two feedback sources, we carried
  out one simulation of the AGN2 scheme for the g51 cluster, in which
  also galactic winds with a velocity $v_w=300\vel$ are included
  (AGN2W hereafter).

Before starting the presentation of the results on the thermal and
enrichment properties of the ICM, we briefly comment on the results
concerning the mass of the central black holes in the simulations
including AGN feedback, and the star formation rate (SFR) history of
the brightest cluster galaxies (BCGs). 

Looking at Table 1, we note that our simulations predict rather large
masses for the super-massive BHs hosted in the central galaxies. Quite
interestingly, the AGN2 runs generally produce BH masses which are
smaller, by a factor 3--5, than for the AGN1 runs. This demonstrates
that including the more efficient ``radio mode'' for the feedback and
distributing the energy in a more uniform way has a significant effect
in regulating gas accretion. Although BCGs are known to host BHs which
are more massive than expected for normal early type galaxies of
comparable mass \citep[e.g.,][]{Lauer2007ApJ...662..808L}, the BH
masses from our simulations seem exceedingly large, also for the AGN2
model. For instance, the BH mass hosted by M87, within the relatively
poor Virgo cluster, is $m_{BH}\simeq (3-4)\times 10^9M_\odot$
\citep[e.g.,][]{Rafferty2006ApJ...652..216R}. This is about a factor
3--5 smaller than found in our AGN2 runs for clusters of comparable
mass, $M_{200}\simeq 2\times 10^{14}\msun$. We note that
  increasing the radio-mode feedback efficiency from $\epsilon_f=0.2$ to
  0.8 (AGN2(0.8) run) reduces the mass of the central black hole in
  the g51 cluster from $m_{BH}\simeq 16.5\times 10^9M_\odot$ to
  $\simeq 6.2 \times 10^9M_\odot$ (see Table 1). This suggests that a
  highly efficient thermalization of the energy extracted from the BH
  is required to regulate gas accretion to the observed level. Quite
  interestingly, we also note that adding the effect of galactic winds
  in the AGN2 scheme (AGN2W run) only provides a marginal reduction of
  the final mass of the central BH. Therefore, although galactic winds
  can play a significant role in regulating star formation within
  relatively small galaxies, they are not efficient in decreasing gas
  density around the largest BHs, so as to suppress their accretion
  rate.

As for the comparison with previous analyses,
\cite{Sijacki2007MNRAS.380..877S} performed a simulation of the same
g676 cluster included in our simulation set at the lower 
resolution, for their feedback scheme based on the injection of AGN
driven bubbles. They found that the final mass of the central BH is
$m_{BH}\simeq 6\times 10^9\msun$. This value is about 30 per cent
lower than what we find for the AGN2 runs of g676. In order to compare
more closely with the result by \cite{Sijacki2007MNRAS.380..877S}, we
repeated the run of the g676 cluster by switching off the
metallicity-dependence of the cooling function. As a result, BH
accretion proceeds in a less efficient way, and the resulting central
BH mass in this case drops to $\simeq 3.5\times 10^9\msun$.

\begin{figure*}
\hbox{
\psfig{figure=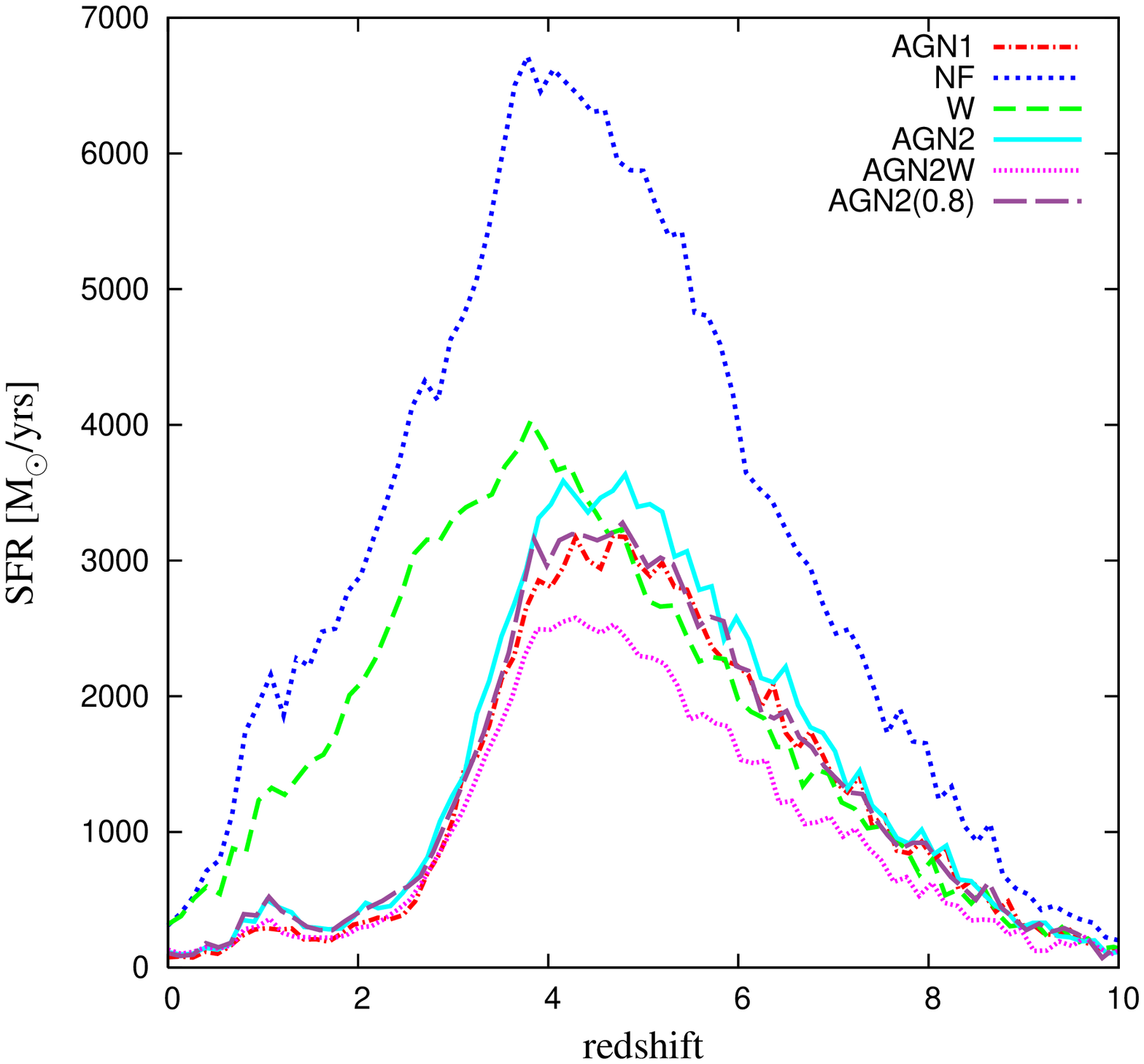,width=8.5cm,angle=-90}
\psfig{figure=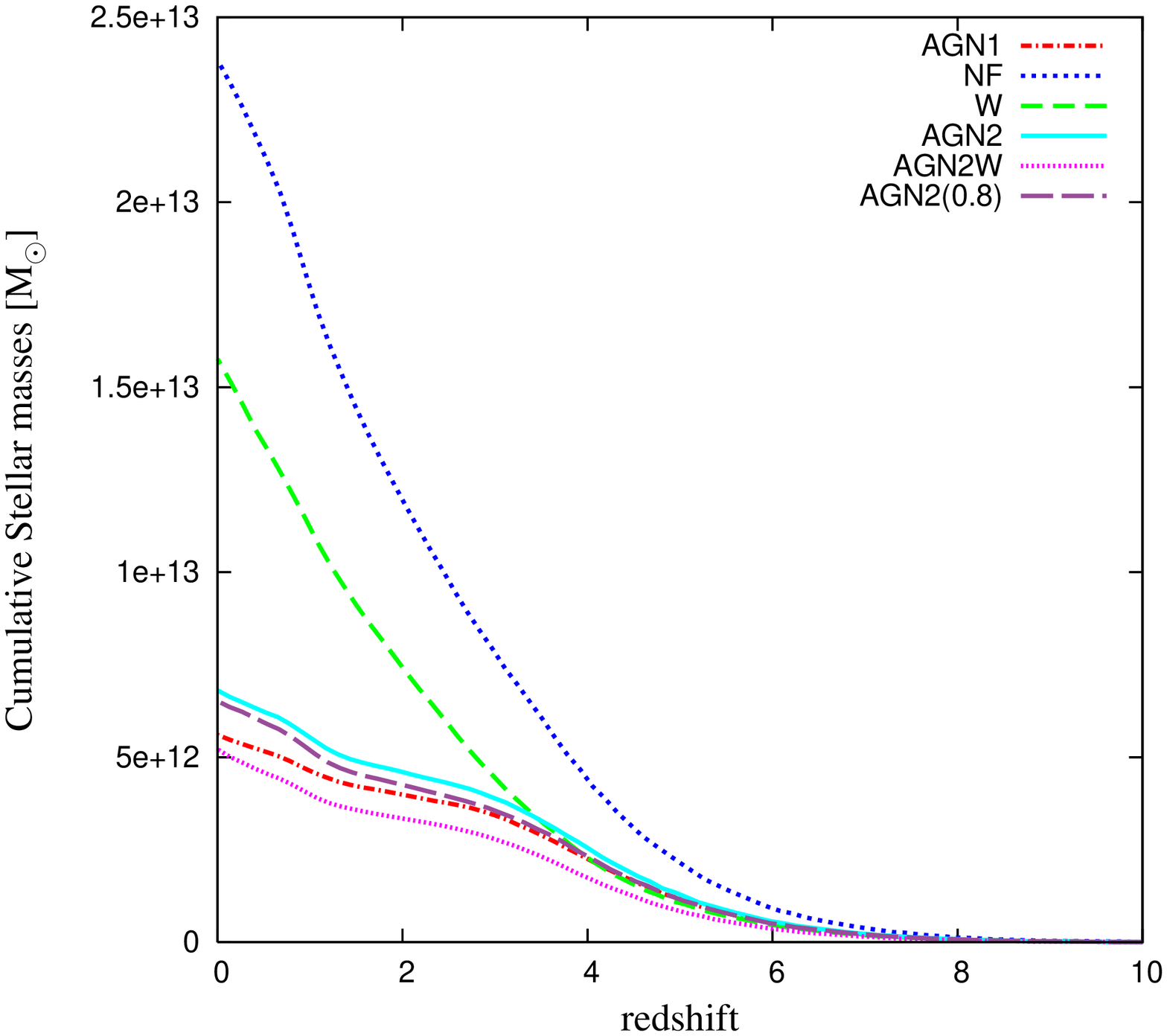,width=8.5cm,angle=-90}
}
\caption{Left panel: the history of star formation rate of the
  brightest cluster galaxy (BCG) in the g51 cluster for the runs with
  no feedback (NF, blue short-dashed), with galactic winds (W, green
  dashed), with standard AGN feedback (AGN1, red dot-dashed) and with
  modified AGN feedback (AGN2, light-blue solid). Also shown are
    the results for the AGN2 run with higher radio-mode BH feedback
    efficiency (AGN2(0.8), purple long-dashed) and for the AGN2 run
    also including galactic winds (AGN2W, magenta dotted). Right
    panel: the stellar mass found in the BCG at $z=0$ that was formed
    before a given redshift.}
\label{Fig:SFR}
\end{figure*}

As for the SFR history of the BCG, this is estimated by identifying
first all the stars belonging to the BCG at $z=0$. This has been
done by running the SKID group-finding algorithm
\citep{Stadel2001PhDT........21S}, using the same procedure described
by \cite{Saro2006MNRAS.373..397S}. After tagging all the star
particles belonging to the BCG, we reconstruct the SFR history by
using the information on the redshift at which each star particle has
been spawned by a parent gas particle, according to the stochastic
algorithm of star formation implemented in the effective model by
\cite{Springel2003MNRAS.339..289S}. The resulting SFR histories for
the BCG of the g51 clusters are shown in the left panel of Figure
  \ref{Fig:SFR}.

As expected, the run with no efficient feedback (NF) produces the
highest star formation at all epochs: SFR has a peak at $z\simeq 4$
and then drops as a consequence of the exhaustion of gas
with short cooling time. Despite this fast reduction of the SFR, its level
at $z=0$ is still rather large, $\simeq 400 \msun$
yr$^{-1}$. As for the run with winds, it has a reduced SFR since the
very beginning, owing to the efficiency of this feedback scheme in
suppressing star formation within small galaxies which form already at
high redshift. Also in this case, the peak of star formation takes
place at $z\sim 4$, but with an amplitude which is about 40 per cent
lower than for the NF runs and with a more gentle decline
afterwards. As for the runs with AGN feedback, their SFR history is
quite similar to that of the W run down to $z\simeq 5$. Star formation
is then suddenly quenched at $z\mincir 4$. We note in general that SFR
for the AGN1 model lies slightly below that of the AGN2 model, as a
consequence of both the different way of distributing energy and, at
low redshift, of the inclusion of the radio mode assumed in the
quiescent BH accretion phase. Suppression of the SF at relatively low
redshift is exactly the welcome effect of AGN feedback. At $z=0$, the
resulting SFR is of about $70 M_\odot$yr$^{-1}$ for both models, a
value which is closer to, although still higher than the typical
values of SFR observed in the BCGs of clusters of comparable mass.
\citep[e.g.,][]{Rafferty2006ApJ...652..216R}. Quite
  interestingly, increasing the feedback efficiency to
  $\epsilon_f=0.8$ in the AGN2(0.8) run does not significantly affect
  the level of low-redshift star formation, while it suppresses star
  formation by only $\sim 10$ per cent around the peak of
  efficiency. Therefore, while a higher efficiency is indeed effective
  in suppressing gas accretion onto the central BH, a further
  reduction of star formation associated to the BCG should require a
  different way of thermalizing the radiated energy. As for the AGN2W
  run, its SFR at high redshift is lower than for the AGN2 run, due to
  the effect of winds, while no significant change is observed at
  $z\mincir 3$.

  In the right panel of Fig.\ref{Fig:SFR} we plot the stellar mass
  found in the BCG at $z=0$ that is formed before a given
  redshift. According to this definition, this quantity is the
  integral of the SFR plotted in the right panel, computed from a
  given redshift $z$ to infinity. This plot clearly shows the
  different effect that winds and AGN feedback have in making the BGC
  stellar population older. In the NF and W runs, the redshift at
  which 50 per cent of the BCG stellar mass was already in place is
  $z_{50}\simeq 2$.  This indicates that even an efficient SN feedback
  is not able to make the stellar population of the BCG older. On the
  contrary, the effect of AGN feedback takes place mostly at
  relatively low reshift. As a consequence the age of the BGC stellar
  population increases, with $z_{50}\simeq 3.0-3.6$, almost
  independent of the detail of the AGN feedback scheme.

\section{Thermal properties of the ICM}
Galaxy clusters are identified in each simulation box by running first
a friends-of-friends (FOF) algorithm over the high resolution DM
particles, using a linking length of $0.15$ in terms of the mean
inter-particle separation. Within each FOF group, we identify the DM
particle having the minimum value of the gravitational potential and
take its position to correspond to the centre of the cluster. All
profiles are then computed starting from this centre. The smallest
radius that we use to compute profiles encompasses a minimum
number of $100$ SPH particles, a criterion that gives numerically
converged results for profiles of gas density and temperature in
non--radiative simulations
\citep[e.g.,][]{Borgani2002MNRAS.336..409B}. As we shall see in the
following, profiles computed with this criterion will extend to
smaller radii for those runs which have a higher gas density at the
centre, while stopping at relatively larger radii for the runs
including AGN feedback, which are characterised by a lower central gas
density.

\subsection{The luminosity-temperature relation}
The relation between bolometric X--ray luminosity, $L_X$ and ICM
temperature, $T$, provided one of the first evidences that
non--gravitational effects determine the thermo-dynamical properties
of the ICM \citep[e.g.,][for a review]{Voit2005RvMP...77..207V}. Its
slope at the scale of clusters is observed to be $L_X\propto T^\alpha$
with $\alpha \simeq 2.5$--3
\citep[e.g.,][]{Horner2001PhDT........88H,Pratt2009A&A...498..361P},
and possibly even steeper or with a larger scatter at the scale of
galaxy groups \citep[e.g.,][]{Osmond2004MNRAS.350.1511O}. These
results are at variance with respect to the prediction, $\alpha =2$,
of self-similar models based only on the effect of gravitational gas
accretion \citep[e.g.,][]{Kaiser1986MNRAS.222..323K}. Attempts to
reproduce the observed $L_X$--$T$ relation with hydrodynamic
simulations of clusters have been pursued by several groups
\citep[see][ for a recent review]{Borgani2008SSRv..134..269B}.
Simulations of galaxy clusters including the effect of star formation
and SN feedback in the form of energy--driven galactic winds produce
results which are close to observations at the scale of clusters,
while generally producing too luminous galaxy groups
\citep[e.g.,][]{Borgani2004MNRAS.348.1078B}. Although a closer
agreement with observations at the scale of groups can be obtained by
using SN--triggered momentum--driven winds
\citep{Dave2008MNRAS.391..110D}, there is a general consensus that
stellar feedback can not reproduce at the same time both the observed
$L_X$--$T$ relation and the low star formation rate observed in
central cluster galaxies. \cite{Puchwein2008ApJ...687L..53P} presented
results on the $L_X$--$T$ relation for simulations of galaxy clusters
which included the bubble--driven AGN feedback scheme introduced by
\cite{Sijacki2007MNRAS.380..877S}. They concluded that, while
simulations not including any efficient feedback (in fact, quite
similar to our NF runs) produce overluminous objects, their mechanism
for AGN feedback is efficient in suppressing the X--ray luminosity of
clusters and groups at the observed level.

\begin{figure*}
\hbox{
\psfig{figure=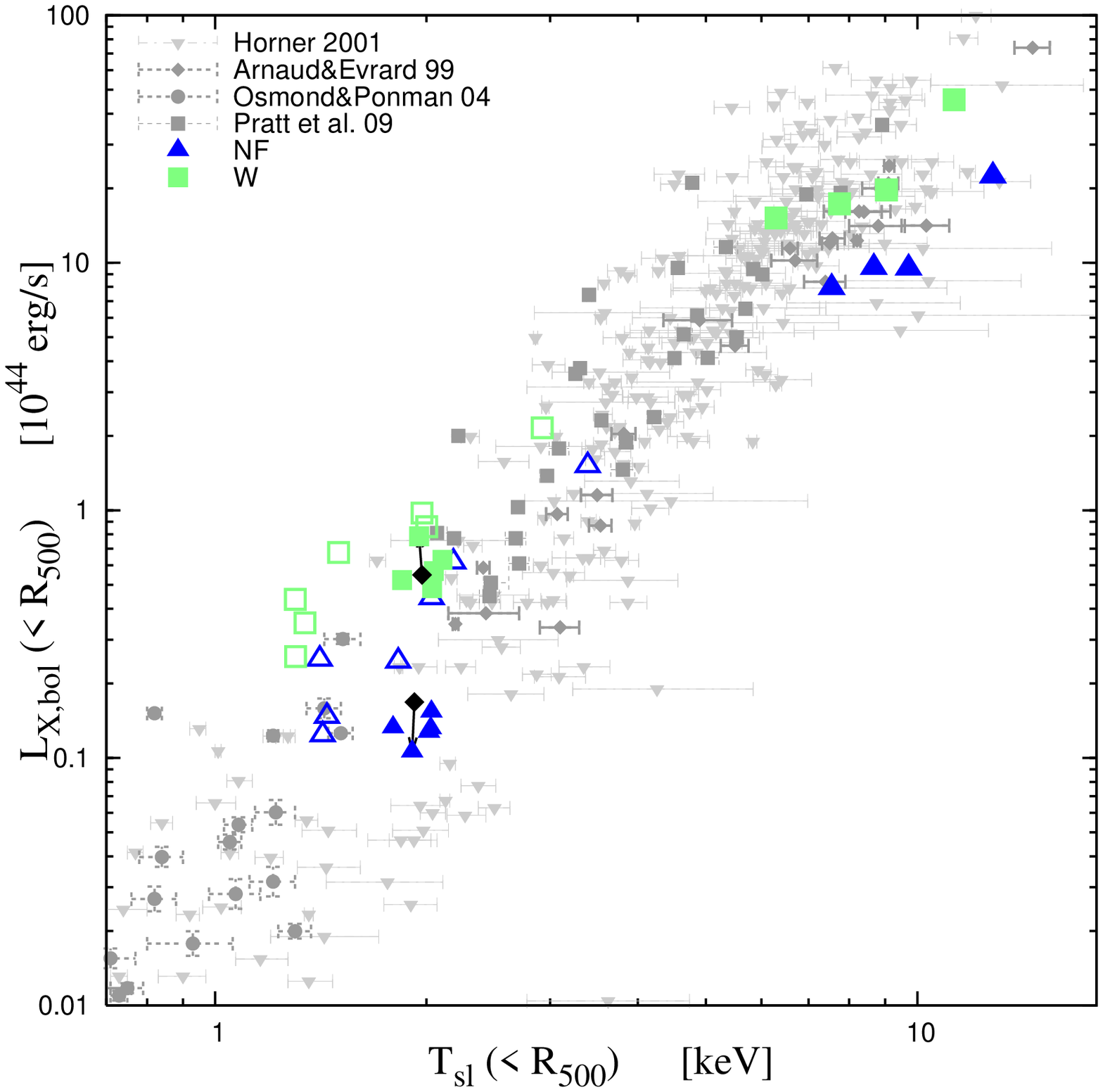,width=9.0cm,angle=-90}
\psfig{figure=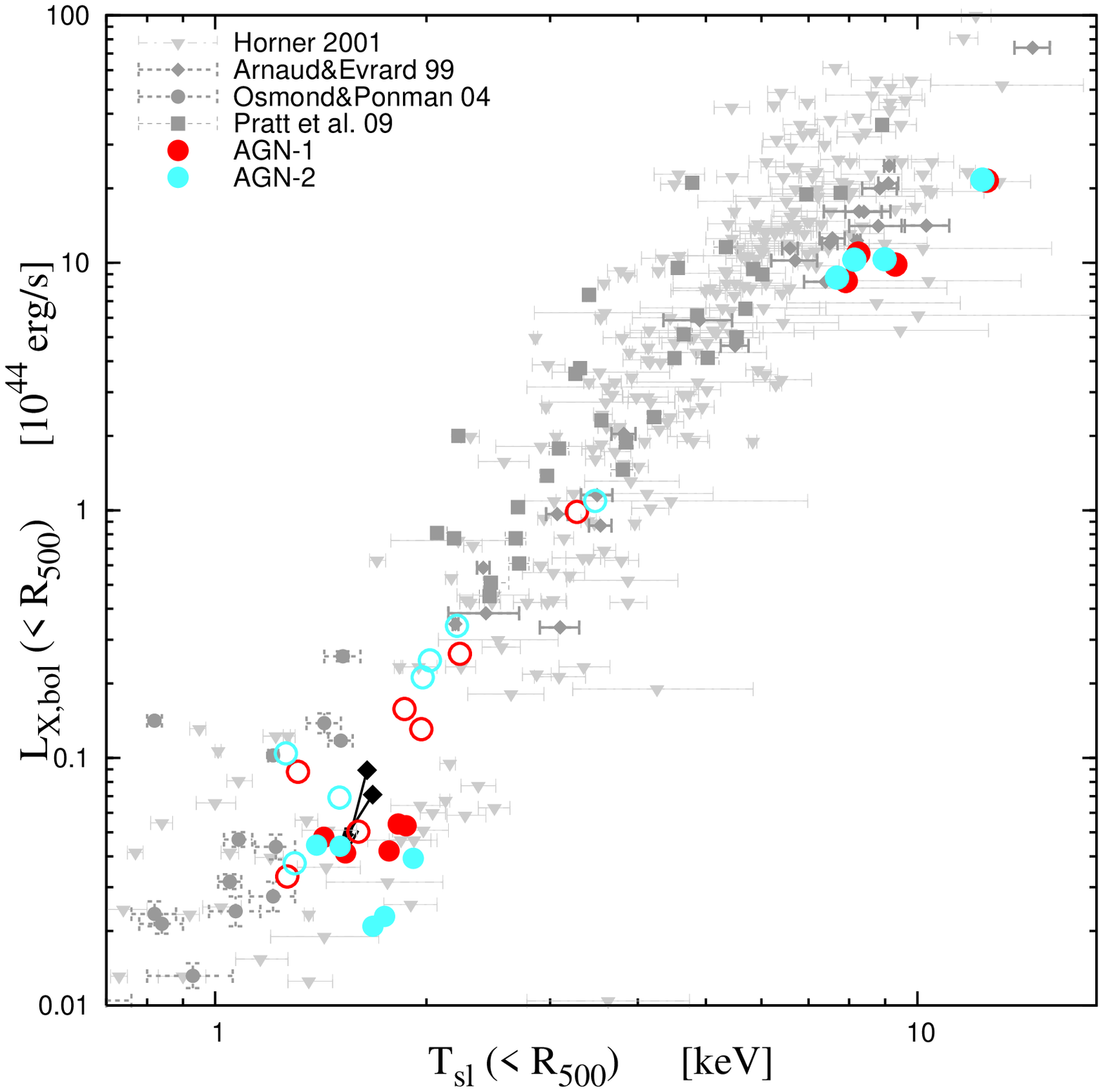,width=9.0cm,angle=-90}
}
\caption{Relation between X--ray luminosity and temperature for
  simulated (coloured symbols) and observed (grey points with
  errorbars) clusters and groups. Observational data points are from
  \protect\cite{Arnaud1999MNRAS.305..631A} (grey diamonds) and 
  \protect\cite{Pratt2009A&A...498..361P} (grey squares) for clusters, from
  \protect\cite{Osmond2004MNRAS.350.1511O} (grey circles) for groups and from
  \protect\cite{Horner2001PhDT........88H} (grey triangles). Data from simulations
  were computed inside $R_{500}$.  Left panel: results for the no
  feedback case (NF, blue triangles) and for the case with galactic
  winds (W, green squares). Right panel: results
  for the runs with standard AGN feedback (AGN1, red circles) and
  with the modified AGN feedback scheme, where also a radio-mode
  regime is included (see text, AGN2; cyan circles). For each series
  of runs, filled and open symbols refer to the main halo within each
  resimulated Lagrangian region and to ``satellite'' halos
  respectively. Black diamonds refer to the runs at $1\times$
  resolution, with arrows pointing to results at $6\times$ higher
  resolution for the g676 cluster.}
\label{Fig:LT}
\end{figure*}

We present here our results on the $L_X$--$T$ relation, keeping in
mind that our simulations differ from those by
\cite{Puchwein2008ApJ...687L..53P} both in the details of the
implementation of the AGN feedback scheme and in the treatment of the
metallicity dependence of the cooling function.  In the left panel of
Figure \ref{Fig:LT} we show the results for our runs based on SN
galactic winds (W runs) and for the runs not including any efficient
feedback (NF runs). Filled symbols refer to the main halo of each
simulated Lagrangian region, while open circles are for the
``satellites''. Although we find several satellites having a
temperature comparable to those of the low-mass main halos (see also
Table 1), we remind that these satellites are described with a mass
resolution six times lower than that of the low-mass main halos. We
note that the NF runs provide a $L_X$--$T$ relation which is not far
from the observed one, especially at the scale of groups. The reason
for this closer agreement, with respect to the result by
\cite{Puchwein2008ApJ...687L..53P} lies in the fact that these authors
adopted a cooling function computed for zero metallicity.
Including the contribution of metal lines to the radiative losses
increases cooling efficiency and, therefore,
gas removal from the hot X--ray emitting phase. However, the price to
pay for this reduction of X--ray luminosity is that a far too large
baryon fraction is converted into stars within clusters (see
below). Quite paradoxically, we also note than including an efficient
feedback in the form of galactic winds turns into an increase of
X--ray luminosity. This is due to the fact that this feedback prevents
a substantial amount of gas from cooling, without displacing it from
the central cluster regions, thus increasing the amount of X-ray
emitting ICM.

In order to asses the effect of the different resolution used for
large and small clusters, we carried out runs of g676 at the same
lower resolution of the massive clusters. The results are shown in
Fig. \ref{Fig:LT} with black diamonds connected with arrows to the
corresponding higher resolution result. We note that resolution
effects go in opposite directions for the NF and W runs. In fact, in
the absence of winds, the runaway of cooling with resolution removes
from the hot phase a larger amount of gas, thus decreasing X--ray
luminosity. On the contrary, higher resolution allows a more accurate
description of kinetic feedback, which starts heating gas at higher
redshift. As a consequence, radiative losses are compensated at higher
resolution for a larger amount of diffuse baryons, which remain in the
hot phase, thereby increasing the X--ray luminosity.

As for the runs with AGN feedback, it is quite efficient in decreasing
X--ray luminosity at the scale of galaxy groups, thus well recovering
the observed $L_X$--$T$ relation.  This conclusion holds for both
implementations of the AGN feedback, which have rather small
differences. Therefore, although the AGN2 scheme is more efficient in
regulating the growth of the central BHs hosted within the BCGs, it
has a marginal effect on the X--ray luminosity. These results are in
line with those found by \cite{Puchwein2008ApJ...687L..53P}, who
however used the injection of heated bubbles to distribute the energy
extracted from the BH accretion. This witnesses that the feedback
energy associated to gas accretion onto super-massive BHs is indeed
able to produce a realistic $L_X$--$T$ relation, almost independent of
the detail of how the energy is thermalised in the surrounding medium.

\subsection{The entropy of the ICM}
Entropy level in central regions of clusters and groups is considered
another fingerprint of the mechanisms which determine the
thermodynamical history of the ICM. Early observational results on the
presence of entropy cores \citep[e.g.,][]{Ponman1999Natur.397..135P}
have been more recently revised, in the light of higher quality data
from Chandra \citep[e.g.,][]{Cavagnolo2009ApJS..182...12C} and
XMM--Newton \citep[e.g.,][]{Johnson2009MNRAS.tmp..469J} observations.
These more recent analyses show that entropy level of clusters and
groups at $R_{500}$ is higher than predicted by non--radiative
simulations, with entropy profiles for relaxed systems continuously
decreasing down to the smallest resolved radii \citep[see
also][]{Sun2009ApJ...693.1142S}.

While radiative simulations of galaxy clusters have been generally
shown to reproduce observationalr results at $R_{500}$
\citep[e.g.,][]{Nagai2007ApJ...668....1N,Dave2008MNRAS.391..110D},
they generally predict too low entropy levels at smaller radii. For
instance, \cite{Borgani2009MNRAS.392L..26B} have shown that the
entropy at $R_{2500}$ can be reproduced by resorting to a fairly
strong pre-heating at $z=4$. However, this pre-heating must target
only relatively overdense regions, to prevent the creation of too
large voids in the structure of the Lyman-$\alpha$ forest at $z\sim 2$
\citep[see also][]{Shang2007ApJ...671..136S}.

\begin{figure*}
\vbox{\hbox{
\psfig{figure=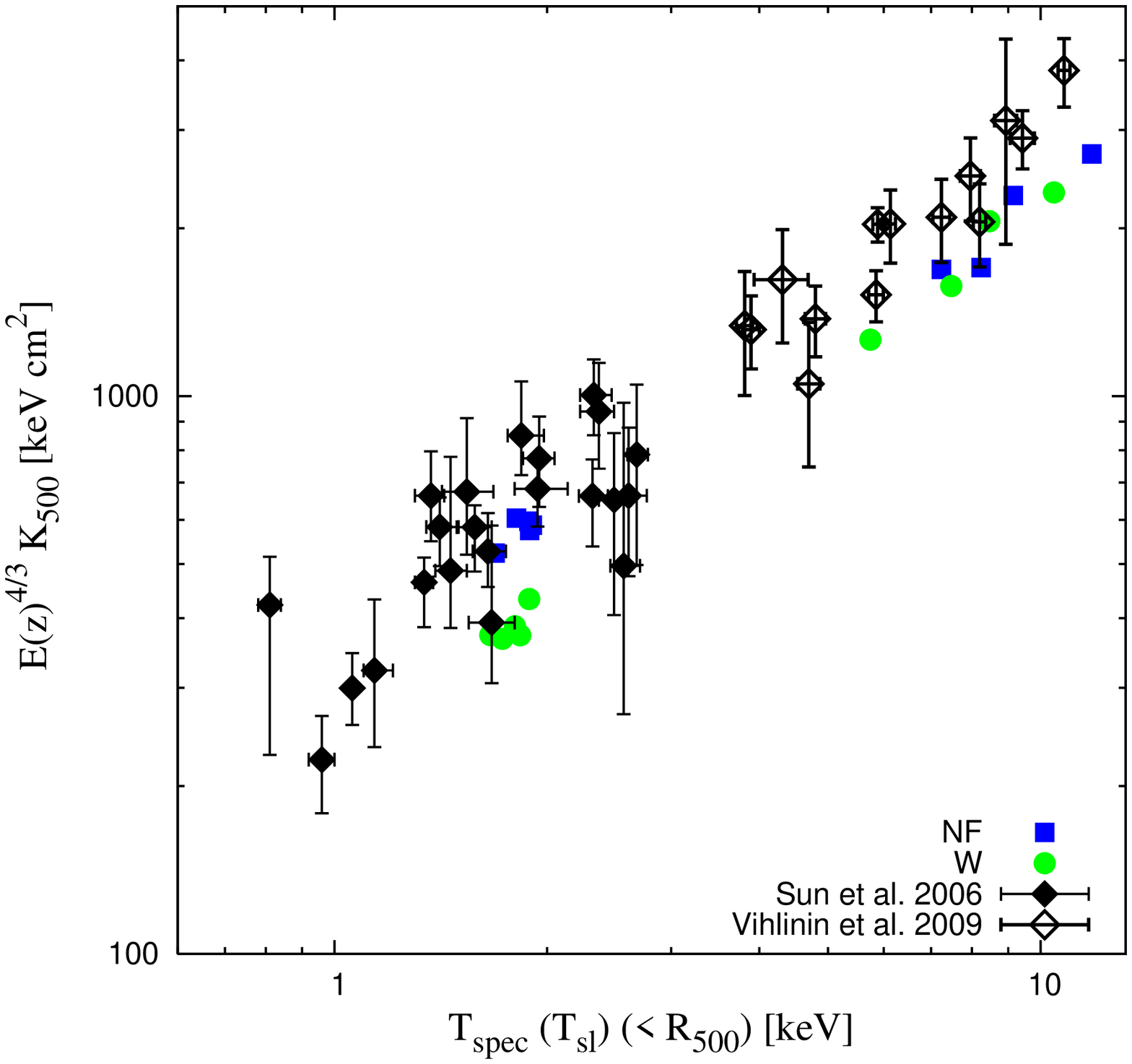,width=9.0cm,angle=-90}
\psfig{figure=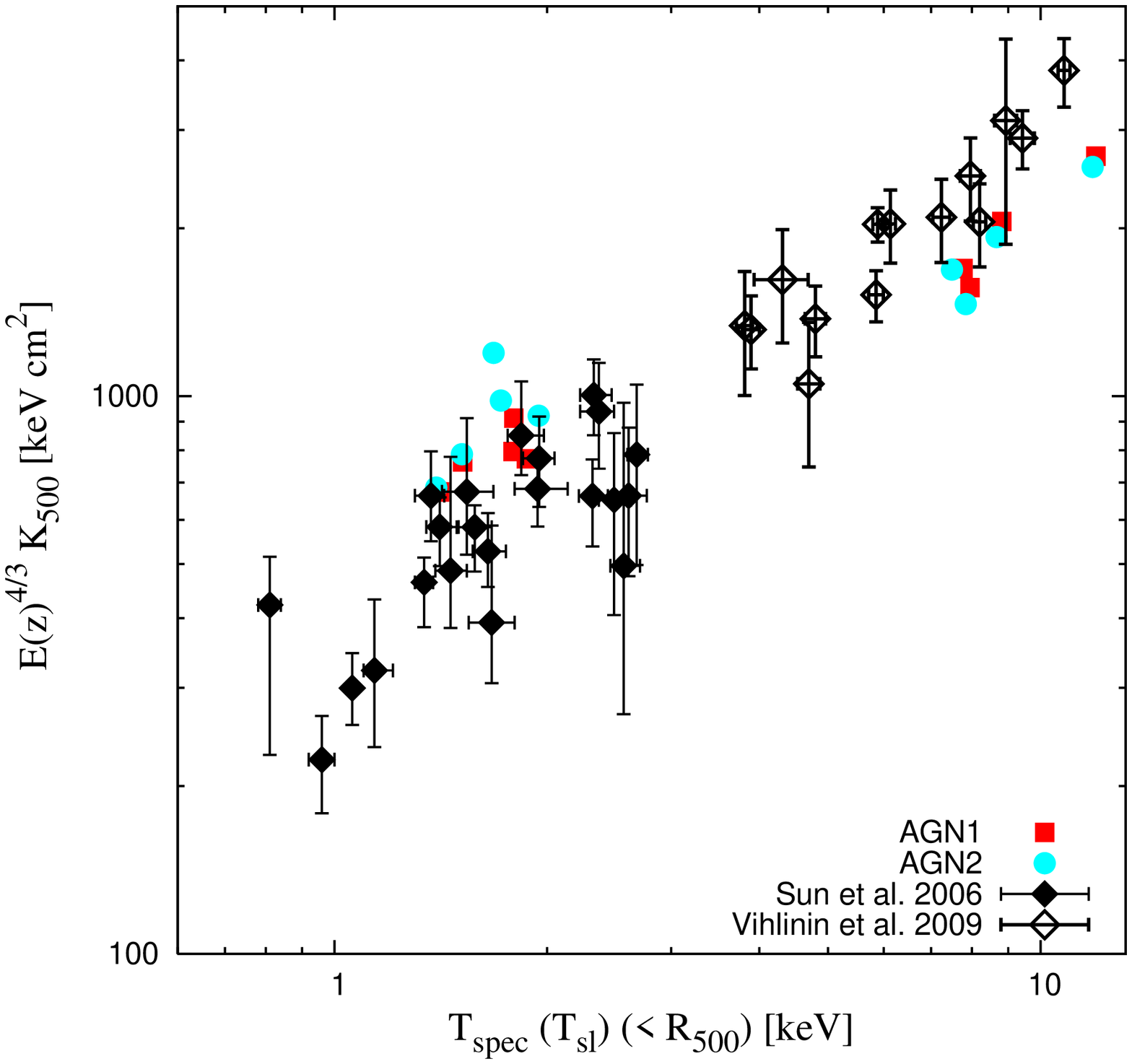,width=9.0cm,angle=-90}
}
\hbox{
\psfig{figure=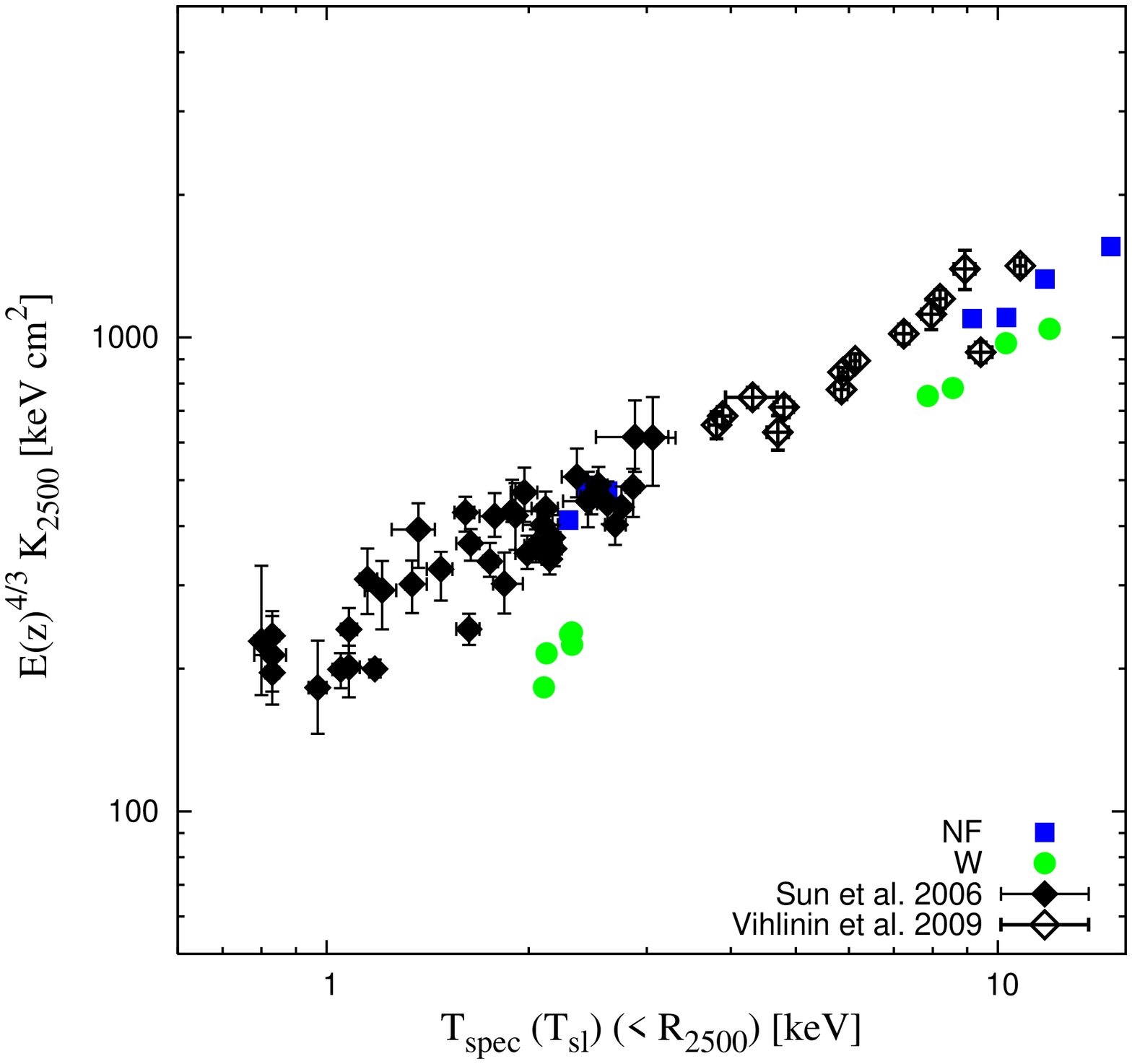,width=9.0cm,angle=-90}
\psfig{figure=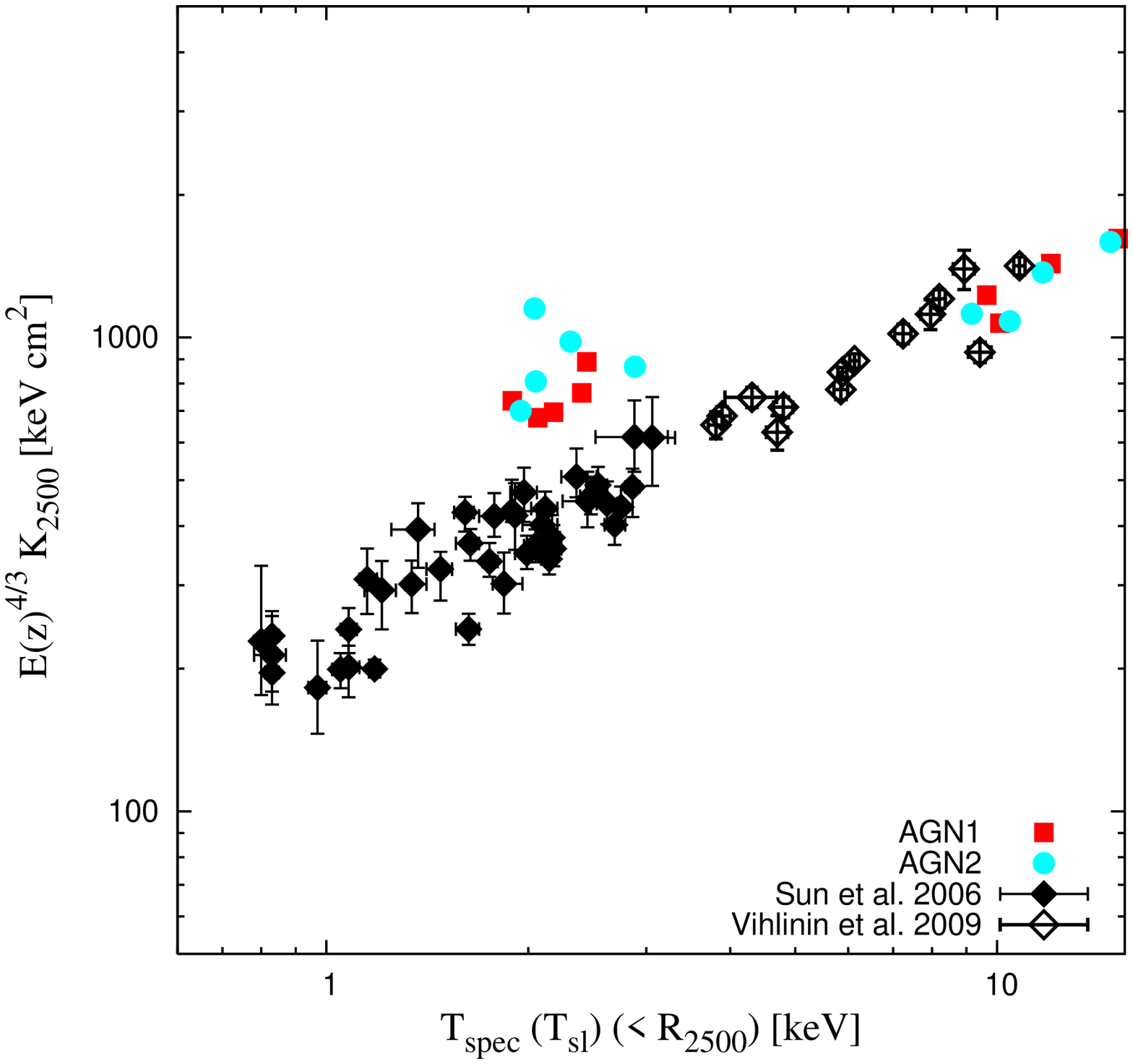,width=9.0cm,angle=-90}
}}
\caption{Relation between entropy and temperature for our simulated
  clusters (coloured circles and squares) and for the observational
  data points at $R_{500}$ (upper panels) and $R_{2500}$ (lower
  panels) from \protect\cite{Sun2009ApJ...693.1142S} (black filled
  diamonds) and \protect\cite{Vikhlinin2009ApJ...692.1033V} (black
  open diamonds). Left and right panels show results for the nine central
  clusters for the runs without (NF: blue squares; W:
  green circles) and with AGN feedback (AGN1: red squares; AGN2:
  cyan circles), respectively.  For a fair comparison with
  observations, spectroscopic-like temperatures of the simulated
  clusters are computed by excluding the regions within
  $0.15R_{500}$.}
\label{Fig:KT}
\end{figure*}

In the following we use the standard definition of entropy, which is
usually adopted in X--ray studies of galaxy clusters
\citep[e.g.,][]{Sun2009ApJ...693.1142S}:
\be
K_\Delta\,=\,{T_{\Delta}\over n_{e,\Delta}^{2/3}\,},
\label{eq:entr}
\ee
where $T_{\Delta}$ and $n_{e,\Delta}$ are the values of gas
temperature and electron number density computed at $R_\Delta$. As for
the temperature, it is computed by following the prescription of
spectroscopic like temperature introduced by
\cite{Mazzotta2004MNRAS.354...10M}.  This definition of temperature
has been shown to accurately reproduce, within few percents, the
actual spectroscopic temperature obtained by fitting spectra of
simulated clusters with a single--temperature plasma model, within the
typical energy bands where detectors on-board of present X--ray
satellites are typically sensitive.

We show in Figure \ref{Fig:KT} the comparison between our simulations
and observational data on groups \citep{Sun2009ApJ...693.1142S} and on
clusters \citep{Vikhlinin2009ApJ...692.1033V} for the relation between
entropy and temperature at $R_{500}$ and $R_{2500}$ (upper and lower
panels, respectively). In order to reproduce the procedure adopted by
\cite{Sun2009ApJ...693.1142S}, we compute the spectroscopic--like
temperature of simulated clusters by excluding the core regions within
$0.15R_{500}$. As for the runs with no efficient feedback (NF) we note
that they produce entropy levels, at both $R_{500}$ and $R_{2500}$,
which are close to the observed ones. This result can be explained in
the same way as that found for the $L_X$--$T$ relation: overcooling,
not balanced by an efficient feedback mechanism, removes a large
amount of gas from the X--ray emitting phase, while leaving in this
hot phase only relatively high entropy gas, which flows in from
larger radii as a consequence of lack of central pressure
support. 

Including winds (W runs) has the effect of increasing the amount of
low-entropy gas, which is now allowed to remain in the hot phase
despite its formally short cooling time, thanks to the continuous
heating provided by winds. As a result, entropy decreases at both
radii, for the same reason for which X--ray luminosity increases. As
for the runs with AGN feedback, its effect is almost negligible for
massive clusters. On the other hand, AGN feedback provides a
significant increase of the entropy level in poor systems, an effect
which is larger at smaller cluster-centric radii. The resulting
entropy is higher than indicated by observational data. Together with
the result on the $L_X$--$T$ relation, this result shows that a rather
tuned energy injection is required, which must be able to suppress
X--ray luminosity by decreasing gas density, while at the same time
reproducing the low entropy level measured at small radii.

\subsection{Temperature profiles}
A number of comparisons between observed and simulated temperature
profiles of galaxy clusters have clearly demomnstrated that a
remarkable agreement exists
at relatively
large radii, $R\magcir 0.2R_{180}$, where the effect of cooling is
relatively unimportant. While this result, which holds almost
independently of the physical processes included in the simulations
\citep[e.g.,][]{Loken2002ApJ...579..571L,Borgani2004MNRAS.348.1078B,
  Kay2007MNRAS.377..317K,Pratt2007A&A...461...71P,Nagai2007ApJ...668....1N},
should be regarded as a success of cosmological simulations
of galaxy clusters, the same simulations have much harder time to
predict realistic profiles within cool-core regions \citep[e.g.,][for
a recent review]{Borgani2008SSRv..134..269B}. In this regime,
radiative simulations systematically produce steep negative
temperature profiles, at variance with observations, as a consequence
of the lack of pressure support caused by overcooling. This further
demonstrates that a suitable feedback mechanisms is required to
pressurise the gas, so as to prevent overcooling and turning the
temperature gradients from negative to positive in the core
regions. While feedback associated to SNe has been proved not to be
successful, AGN feedback is generally considered as a likely solution
for simulations to produce realistic cool cores. Based on simulations
of one relatively low--mass cluster, \cite{Sijacki2008MNRAS.387.1403S}
found that AGN feedback can provide reasonable temperature profiles
only if a population of relativistic particles is injected along with
thermal energy in inflated bubbles.

\begin{figure*}
\hbox{
\psfig{figure=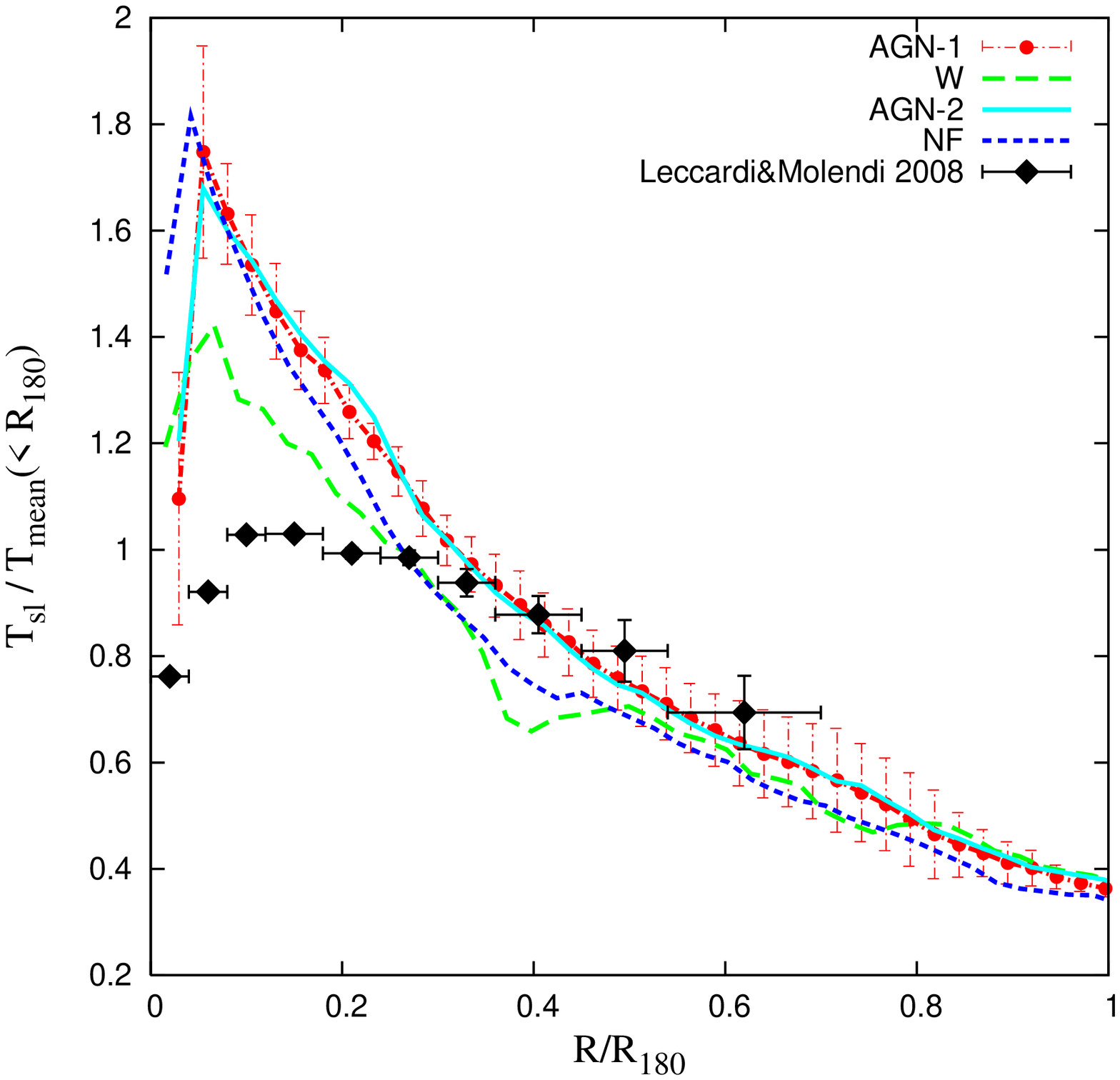,width=9.0cm,angle=-90}
\psfig{figure=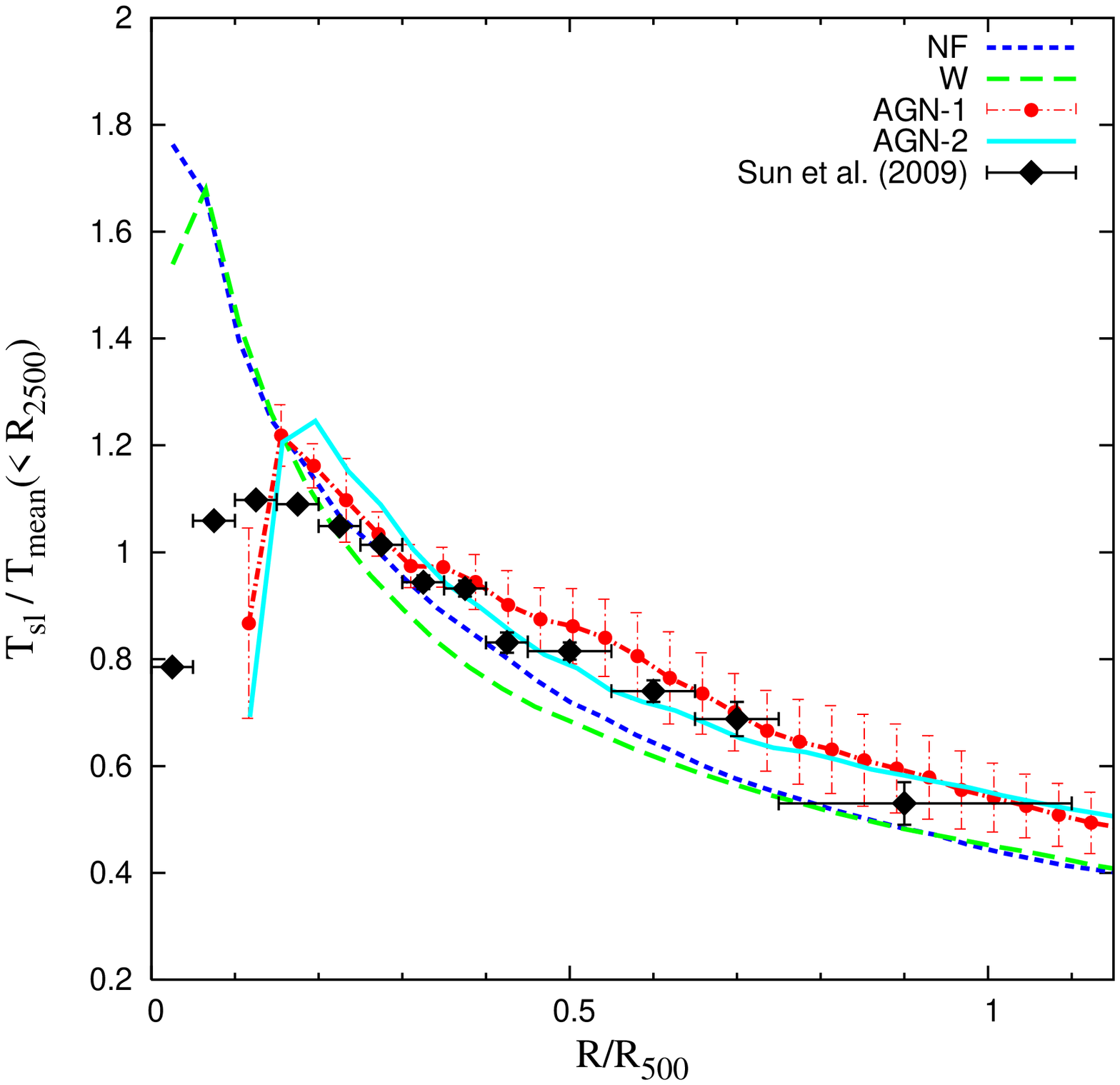,width=9.0cm,angle=-90}
}
\caption{Comparison between the temperature profiles for simulated
  and observed clusters with $T\magcir 3$ keV (left panel) and for
  groups with $T\mincir 3$ keV (right panel). In each panel, different lines
  corresponds to the average simulated profiles computed over the four
  main massive clusters in the left panel and over the five main
  low-mass clusters in the right panel, for the different sets of
  runs: no feedback (NF, blue short dashed), galactic winds (W, green long
  dashed), standard AGN feedback (AGN1, red dot-dashed), modified AGN
  feedback (AGN2, cyan solid). For reasons of clarity, we show with
  errorbars the r.m.s. scatter over the ensemble of simulated clusters
  only for the AGN1 runs. Observational data points for clusters in
  the left panel are taken from \protect\cite{Leccardi2008A&A...486..359L},
  while those for groups in the right panel are from
  \protect\cite{Sun2009ApJ...693.1142S}.}
\label{Fig:Tprof}
\end{figure*}

We present in Figure \ref{Fig:Tprof} the comparison between simulated
and observed temperature profiles for galaxy clusters with $T\magcir
3$ keV (left panel) and for poorer clusters and groups with $T\mincir
3$ keV (right panel). Observational results are taken from
\cite{Leccardi2008A&A...486..359L} and \cite{Sun2009ApJ...693.1142S}
for rich and poor systems,
respectively. As for rich clusters, none of the implemented feedback
scheme is capable to prevent the temperature spike at small radii,
while all models provide a temperature profile quite similar to
the observed one at $R\magcir 0.3 R_{180}$. The situation is different
for groups. In this case, both schemes of AGN feedback provide results
which go in the right direction. While galactic winds are 
not able to significantly change the steep negative temperature gradients,
the AGN1 and the AGN2 feedback schemes pressurise the ICM in the
central regions, thus preventing adiabatic compression in inflowing
gas. From the one hand, this result confirms that a feedback scheme
not related to star formation goes indeed in the right direction of
regulating the thermal properties of the ICM in cool core regions. On
the other hand, it also demonstrates that the schemes of AGN feedback 
implemented in our simulations do an excellent job at the scale of
galaxy groups, while they are not efficient enough at the scale of
massive clusters.

\subsection{The gas and star mass fractions}

\begin{figure*}
\hbox{
\psfig{figure=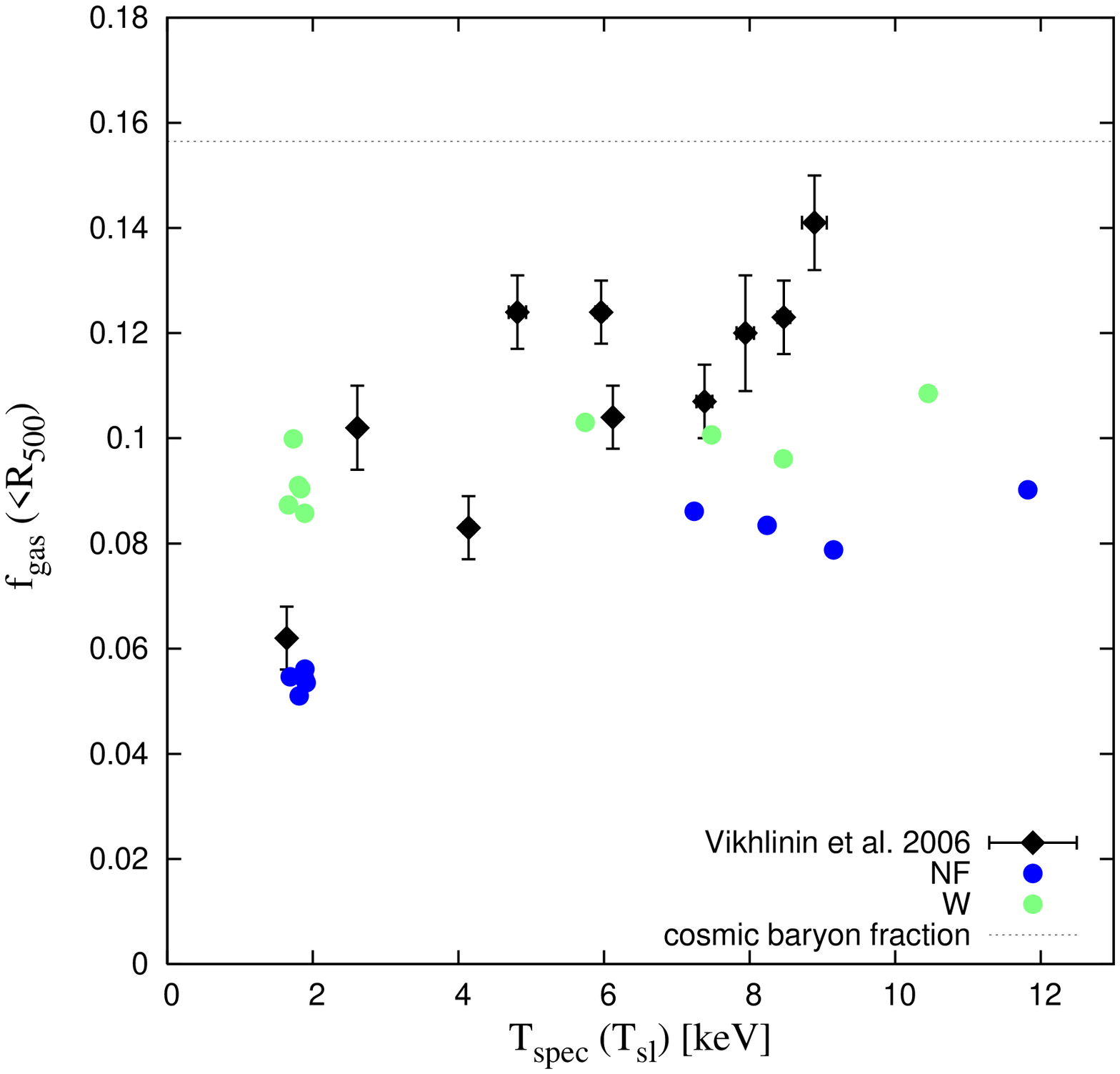,width=9.0cm,angle=-90}
\psfig{figure=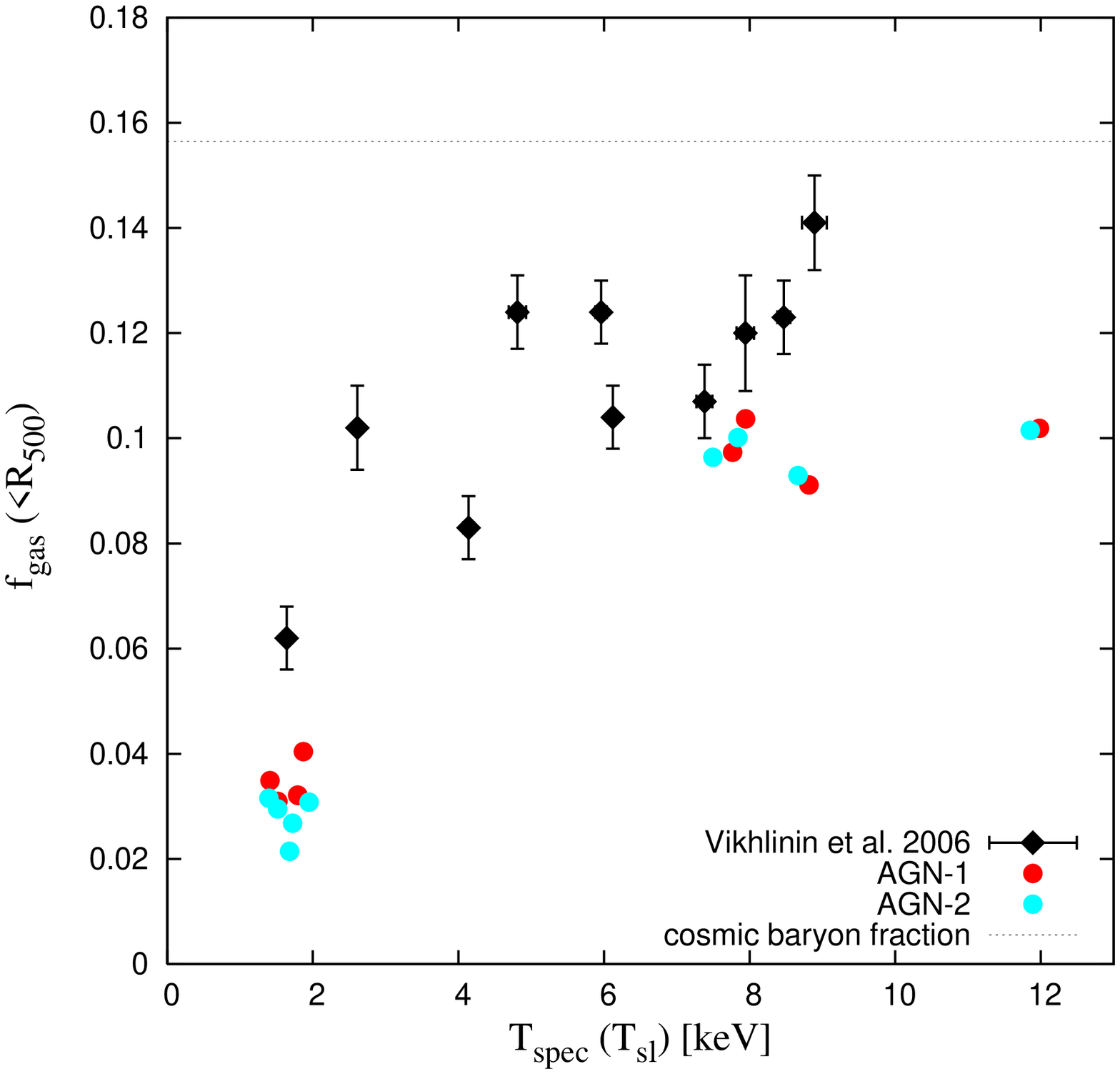,width=9.0cm,angle=-90}
}
\caption{Comparison of the gas fraction within $R_{500}$ in
  simulations (coloured circles) and observational data from Chandra
  data (diamonds with errorbars) analysed by
  \protect\cite{Vikhlinin2006ApJ...640..691V}. Left panel: results for
  the no feedback (NF) runs (dark blue) and for the runs with galactic
  winds (W, light green). Right panel: results for the runs with the
  standard AGN feedback (AGN1, dark red) and with modified AGN
  feedback (AGN2, light cyan). The horizontal dotted line marks the
  cosmic baryon fraction assumed in the simulations.}
\label{Fig:fgas}
\end{figure*}

\begin{figure*}
\hbox{
\psfig{figure=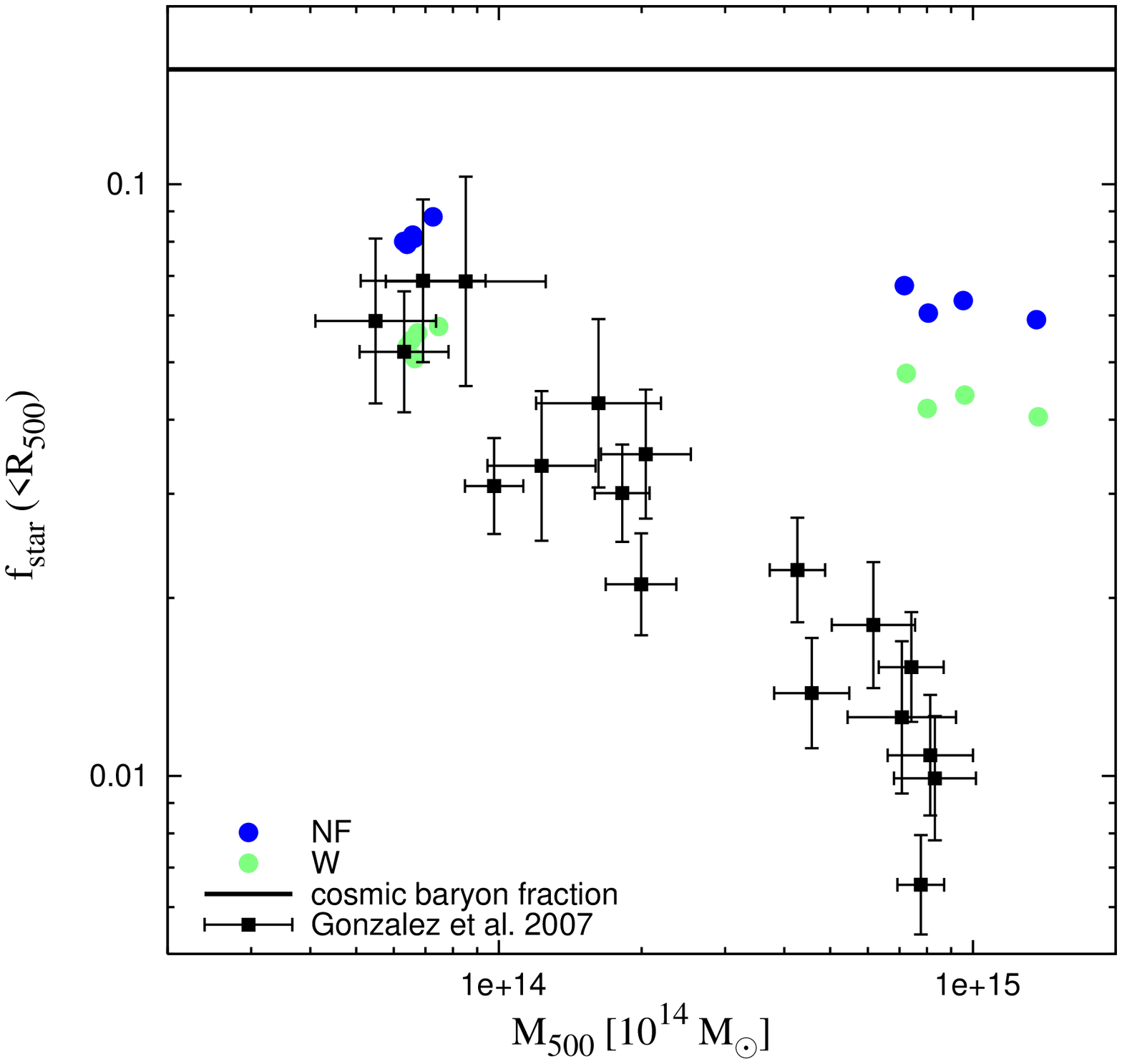,width=9.0cm,angle=-90}
\psfig{figure=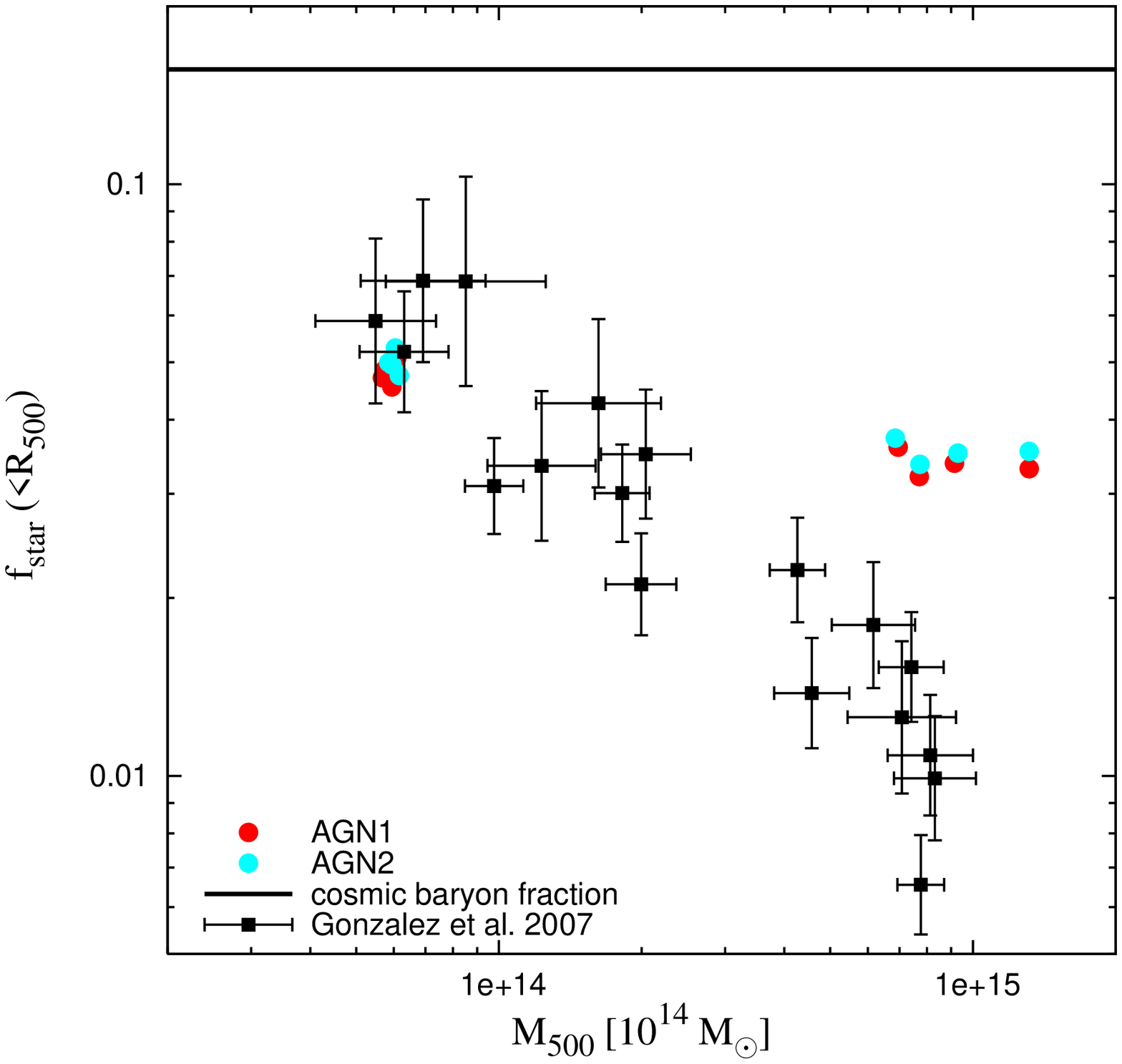,width=9.0cm,angle=-90}
}
\caption{Comparison of the star fraction within $R_{500}$ in
  simulations (coloured circles) and observations (squares with
  errorbars). Left panel: results for the no feedback (NF) runs (dark
  blue) and for the runs with galactic winds (W, light green). Right
  panel: results for the runs with the standard AGN feedback (AGN1,
  dark red) and with modified AGN feedback (AGN2, light cyan).
  Observational points are from
  \protect\cite{Gonzalez2007ApJ...666..147G} where stellar mass
  includes the brightest cluster galaxy (BCG), intra-cluster light
  (ICL) and galaxies within $R_{500}$.}
\label{Fig:fstar}
\end{figure*}

The inventory of baryons within galaxy clusters represents an
important test to understand both the efficiency of star formation and
how the gas content is affected by feedback mechanisms. In general,
the difficulty of regulating gas cooling in cluster simulations causes
a too large stellar mass fraction
\citep[e.g.,][]{Borgani2004MNRAS.348.1078B,Kay2007MNRAS.377..317K,
  Nagai2007ApJ...668....1N,Dave2008MNRAS.391..110D}, which in turn
should correspond to a too low fraction of gas in the diffuse ICM.

We show in Figure \ref{Fig:fgas} the comparison between simulation
results and observational data on the mass fraction of hot gas as a
function of temperature from \cite{Vikhlinin2006ApJ...640..691V}.  Gas
in simulated clusters is assigned to the hot phase if it is not
associated to multi-phase gas particles and if its temperature exceeds
$3\times 10^4$K. A comparison between the runs with no feedback (NF)
and with galactic winds (W) shows that the latter are characterised in
general by larger $f_{gas}$ values. This is in line with the results
on the $L_X$--$T$ relation and confirms that winds are effective in
suppressing cooling, thus increasing the hot baryon fraction. While
there is a reasonable agreement at the scale of poor clusters,
simulations show a weak trend with temperature, with $f_{gas}$
for the hotter systems having values well below the observed ones. As
for AGN feedback, it has the effect of increasing $f_{gas}$ for the
most massive clusters, although the resulting gas fraction is still
below the observational level by about 30 per cent. On the contrary,
at the scale of poor clusters the effect of AGN feedback is that of
decreasing $f_{gas}$ below the observational limit. Indeed, while in
rich systems the effect of AGN feedback is that of reducing
overcooling, thereby leaving a larger amount of gas in the hot phase,
in poor systems it is so efficient as to displace a large amount of
gas outside the cluster potential wells.

Our result on the low value of $f_{gas}$ at the scale of rich
clusters, even in the presence of AGN feedback, is in line with the
somewhat low value of $L_X$ seen in Fig. \ref{Fig:LT}. However, this
result is in disagreement with that presented by
\cite{Puchwein2008ApJ...687L..53P}, who showed instead a good
agreement at all temperatures between their simulations including AGN
feedback and observational data. There may be two reasons for this
difference. Firstly, the scheme to inject AGN-driven high-entropy
bubbles used by \cite{Puchwein2008ApJ...687L..53P} could provide a
more efficient means of stopping cooling in central cluster regions,
at the same time preventing excessive gas removal in low-mass
systems. Secondly, unlike \cite{Puchwein2008ApJ...687L..53P} we
include the dependence of metallicity in the cooling function. As
already discussed, this significantly enhances cooling efficiency and,
therefore, the removal of gas from the hot phase. In order to verify
the impact of this effect, we repeated the AGN2 run of g51, by
assuming zero metallicity in the computation of the cooling
function. As a result, we find that $f_{gas}$ increases from 0.09 to
010. From the one hand, this result implies that the more efficient
gas accretion onto BHs, due to metal-cooling, provides a stronger
energy feedback which, in turn, balances the higher cooling
efficiency. On the other hand, it also implies that the main reason
for the difference with respect to \cite{Puchwein2008ApJ...687L..53P}
should be rather ascribed to the different way in which energy
associated to BH accretion is thermalised in the surrounding medium.

As for the behaviour of the mass fraction in stars, we show in Figure
\ref{Fig:fstar} the comparison between our simulations and
observational results from \cite{Gonzalez2007ApJ...666..147G}, who
also included in the stellar budget the contribution from diffuse
intra-cluster stars \citep[see also][]{Giodini2009arXiv0904.0448G}.
These results confirm that none of our simulations are able to
reproduce the observed decrease of $f_{star}$ with increasing
temperature.  Overcooling is indeed partially prevented in the
presence of winds and, even more, with AGN feedback. However, while
simulation results for poor clusters are rather close to observations,
overcooling in massive clusters is only partially alleviated by AGN
feedback, with values of $f_{star}$ which are larger than
the observed ones by a factor 2--3.

\vspace{0.3truecm} In summary, the results presented in this section
demonstrated that AGN feedback has indeed a significant effect in
bringing the $L_X$--$T$ relation and the entropy level of the ICM
closer to observational results, while regulating cooling in the
central regions. However, the effect is not yet large enough to
produce the correct temperature structure in the cool core of massive
clusters and, correspondingly, the correct share of baryons between
the stellar and the hot gas phase.

\section{Metal enrichment of the ICM}
The X-ray spectroscopic studies of the content and distribution of
metals in the intra-cluster plasma provides important information on
the connection between the process of star formation, taking place on
small scales within galaxies, and the processes which determine the
thermal properties of the ICM. The former affects the quantity of
metals that are produced by different stellar populations, while the
latter gives us insights on gas-dynamical processes, related both to
the gravitational assembly of clusters and to the feedback mechanisms
that displace metal-enriched gas from star forming regions.

The detailed model of chemical evolution included in the \gadget code
by T07 allows us to follow the production of heavy elements and to
study how their distribution is affected by the adopted feedback
schemes. The analysis presented in this section is aimed at
quantifying the different effects that galactic outflows triggered by
SN explosions and AGN feedback have on the enrichment pattern of the
ICM. We present results on the Fe distribution and the
corresponding abundance profiles, the enrichment age within clusters
and groups, the relation between global metallicity and ICM
temperature, and the relative abundance of Si with respect to
Fe. Results from simulations will be compared to the most recent
observational data from the Chandra, XMM-Newton and Suzaku
satellites. All the abundance values will be scaled to the the solar
abundances provided by \cite{Grevesse1998SSRv...85..161G}.

\begin{figure*}
\hbox{
\hspace{1.5truecm}
\psfig{figure=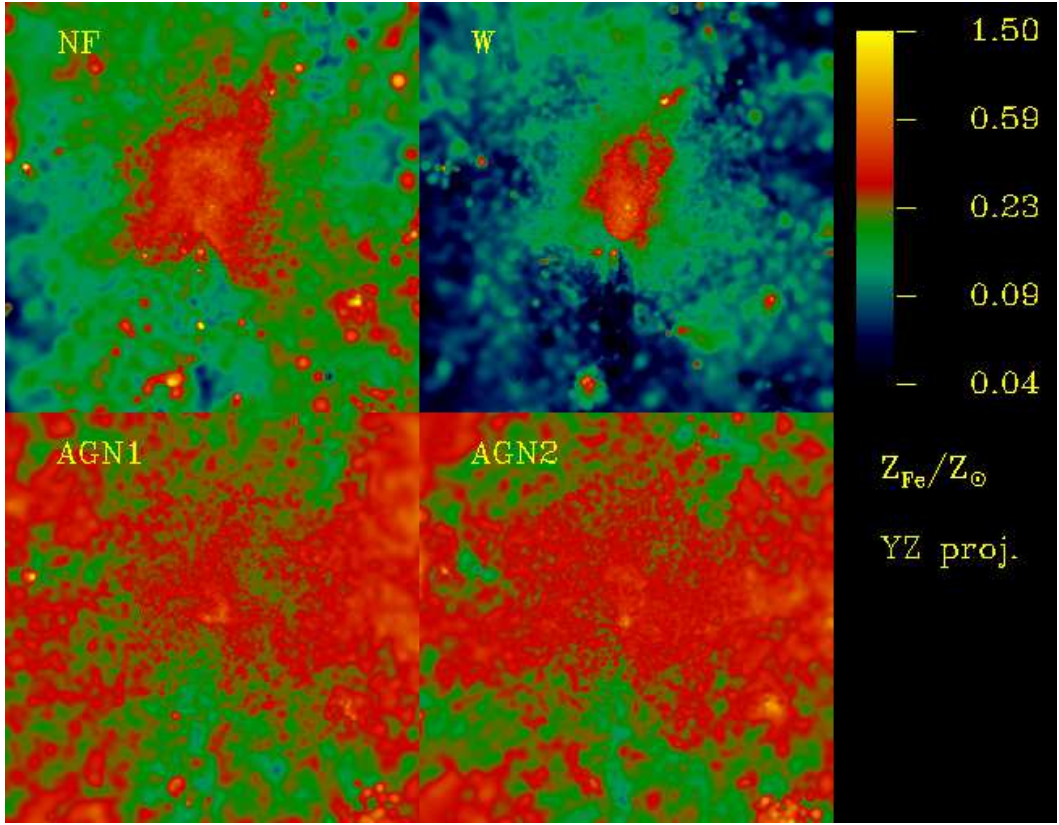,width=14.0cm,angle=0}
}
\caption{Maps of emission weighted Fe abundance in the g51 cluster for
  the runs without feedback (NF, top left), with winds (W, top right)
  and with AGNs (AGN1 and AGN2, bottom left and bottom right,
  respectively). Each map has a side of $2 R_{vir}$. Abundance values
  are expressed in units of the solar value, as reported by
  \protect\cite{Grevesse1998SSRv...85..161G}, with color coding
  specified in the right bar.}
\label{Fig:mapsg51}
\end{figure*}

To qualitatively appreciate the effect of different feedback
mechanisms on the ICM enrichment pattern, we show in Figure
\ref{Fig:mapsg51} the maps of the emission--weighted Fe abundance for
the different runs of the g51 massive cluster. Here and in the
following, we will rely on emission--weighted estimates of metal
abundances. \cite{Rasia2008ApJ...674..728R} have shown that
this emission--weighted estimator actually reproduces quite closely
the values obtained by fitting the X--ray spectra of simulated
clusters, for both Iron and Silicon. As for Oxygen, the
emission--weighted estimator has been shown to seriously overestimate
the corresponding abundance, especially for hot ($T\magcir 3$ keV)
systems.

In the run with no AGN feedback (upper panels of
Fig.\ref{Fig:mapsg51}), we clearly note that including galactic winds
produces a level of enrichment which is lower than that in the NF run
outside the core region, as a consequence of the lower level of star
formation. Therefore, although galactic ejecta are known to be rather
efficient in spreading metals in the intergalactic medium at high
redshift ($z\magcir 2$; e.g.,
\citealt{Oppenheimer2008MNRAS.387..577O,Tescari2009MNRAS.397..411T},
Tornatore et al. 2009, in preparation), their effect is not strong
enough to compensate the reduction of star formation within cluster
regions. In the NF run we note the presence of highly enriched gas
clumps which coincide with the halos of galaxies where intense star
formation takes place.

A rather different enrichment pattern is provided by AGN feedback
(lower panels of Fig.\ref{Fig:mapsg51}).  Despite the total amount of
stars produced in these two runs is smaller than for the run including
galactic winds, AGN feedback is highly efficient in spreading metals
at high redshift, mostly in correspondence of the peak of BH accretion
activity. This demonstrates that AGN feedback provides a rather high
level of diffuse enrichment in the outskirts of galaxy clusters and in
the inter-galactic medium surrounding them at low redshift.

\subsection{Profiles of Iron abundance}

\begin{figure*}
\hbox{
\psfig{figure=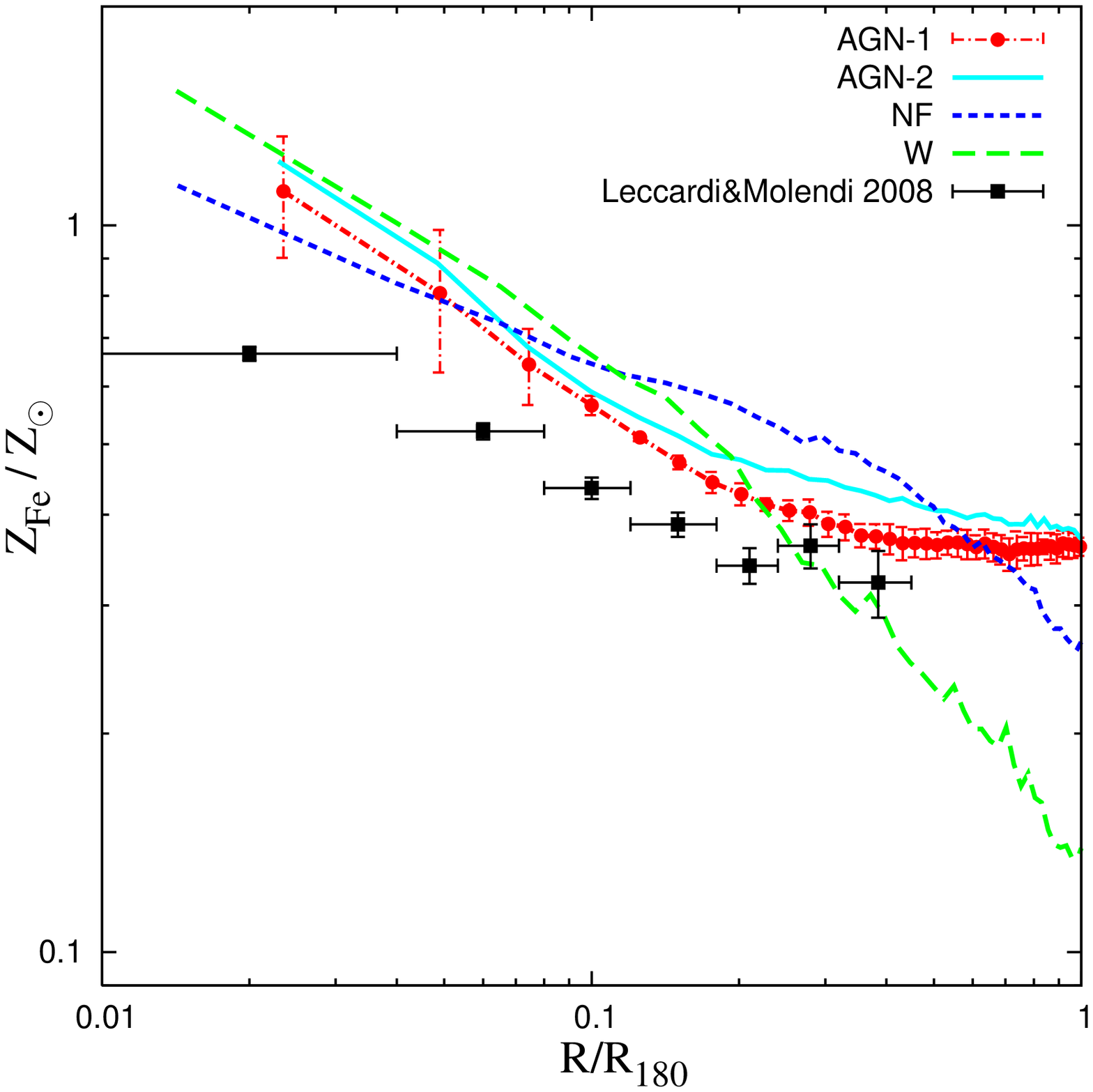,width=9.0cm,angle=-90}
\psfig{figure=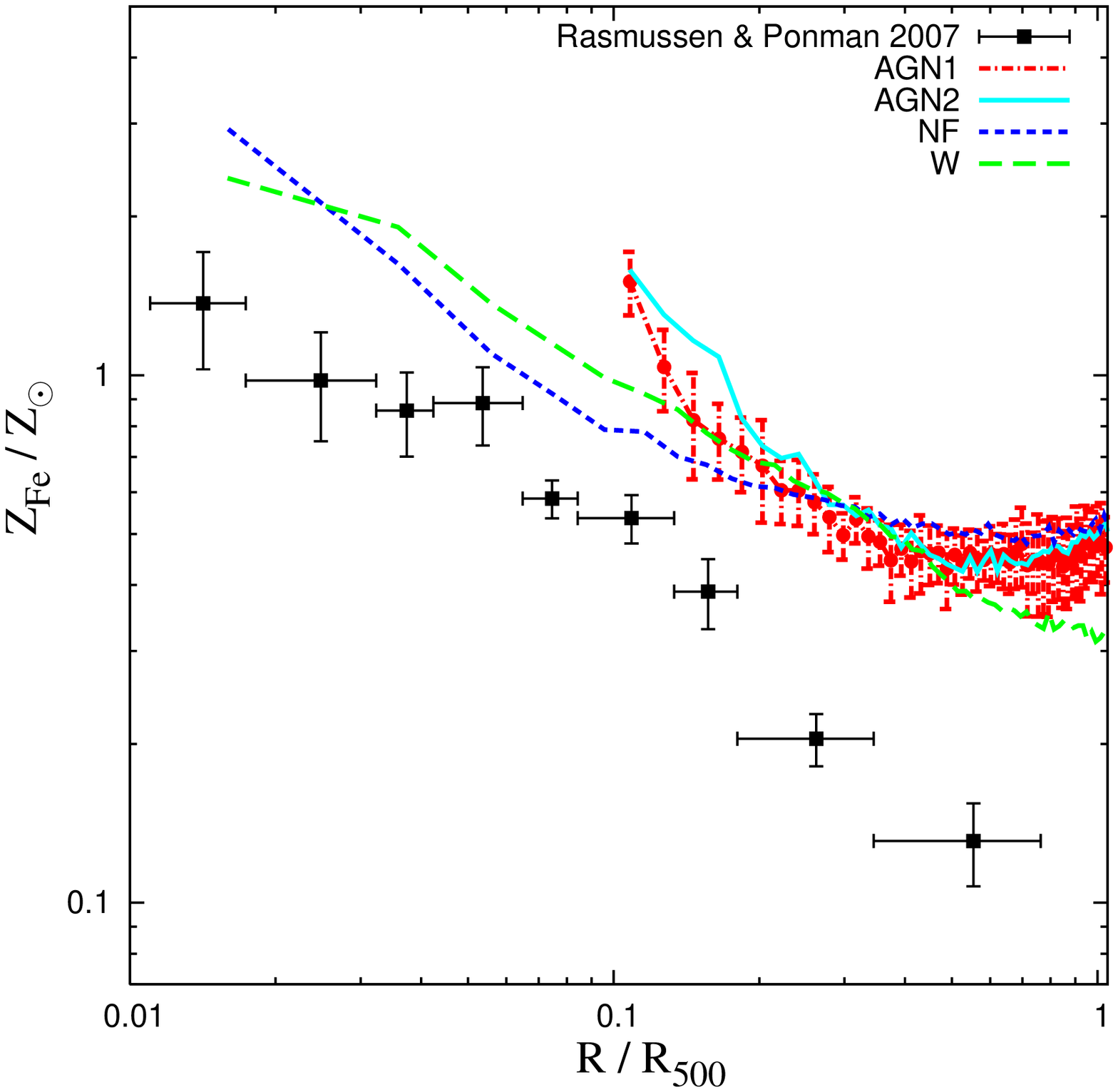,width=9.0cm,angle=-90}
}
\caption{Comparison between the observed and the simulated profiles of
  emission-weighted Iron metallicity. Left panel: average \Zfe
  profiles for galaxy clusters with $T_{500}>3$ keV. Observational
  data points are taken from
  \protect\cite{Leccardi2008A&A...487..461L}. Right panel: average
  \Zfe profiles for the five simulated galaxy groups with $T_{500}<3$
  keV. Observational data points are taken from
  \protect\cite{Rasmussen2007MNRAS.380.1554R}.  In both panels
  different lines correspond to the average profiles computed for the
  different runs: no feedback (NF, blue short dashed), galactic winds
  (W, green long dashed), standard AGN feedback (AGN1, red
  dot-dashed), modified AGN feedback (AGN2, cyan solid). For reasons
  of clarity, we show with $1 \sigma$ errorbars over the ensemble of
  simulated clusters only for the AGN1 runs.}
\label{Fig:met_profs}
\end{figure*}

We show in Figure \ref{Fig:met_profs} the emission-weighted Iron
abundance profiles obtained by averaging over the four simulated
clusters with $T_{sl} > 3$ keV (left panel) and the five galaxy groups
with $T_{sl} < 3$ keV (right panel), compared with observational
results. Each panel reports the results for the four adopted feedback
schemes. For reasons of clarity we report the $1 \sigma$ scatter
computed over the ensemble of simulated clusters only for the AGN1
runs.

As for rich clusters, simulation predictions are compared with the
observational results by \cite{Leccardi2008A&A...487..461L}. These
authors analysed about 50 clusters with $T \magcir 3$ keV, that were
selected from the XMM-Newton archive in the redshift range $0.1 \leq z
\leq 0.3$. After carrying out a detailed modelling of the background
emission, they recovered metallicity profiles out to $\simeq 0.4
R_{180}$. The results of this analysis show a central peak of
$Z_{Fe}$, followed by a decline out to $0.2 R_{180}$, while beyond
that radius profiles are consistent with being flat, with
$Z_{Fe}\simeq 0.3 Z_{Fe,\odot}$ using the solar abundance value by
\cite{Grevesse1998SSRv...85..161G} ($\simeq 0.2$ in units of the solar
abundance by \cite{Anders1989GeCoA..53..197A}, as reported by
\citealt{Leccardi2008A&A...487..461L}).

All our simulations predict the presence of abundance gradients in the
central regions, whose shape is in reasonable agreement with the
observed one, at least for $R\mincir 0.1R_{180}$. The lowest
enrichment level is actually found for the NF run, despite the fact
that this model produces the most massive BCGs. The reason for this
lies in the highly efficient cooling that selectively removes the most
enriched gas, which has the shortest cooling time, thus leaving metal
poorer gas in the diffuse phase. Runs with galactic winds (W) and AGN
feedback (AGN1 and AGN2) are instead able to better regulate gas
cooling in central region, thus allowing more metal-rich gas to
survive in the hot phase.  For this reason, W and AGN runs predict
profiles of \Zfe which are steeper than for the NF runs in the central
regions, $\mincir 0.1 R_{180}$. Quite interestingly, the effect that
different feedback mechanisms have in displacing enriched gas and
regulating star formation almost balance each other in the central
cluster regions, thus producing similar profiles. However, the
different nature of SN-powered winds and AGN feedback leaves a clear
imprint at larger radii.

As for the runs with no feedback (NF), they produce a rather high
level of enrichment out to $\sim 0.3R_{180}$, while rapidly declining
at larger radii. In this model, the high level of star formation
provides a strong enrichment of the gas in the halo of galaxies which
will merge in the clusters. During merging, this gas is ram--pressure
stripped, thus contributing to enhance the enrichment level of the
ICM. The situation is different for the runs with winds. As already
mentioned, galactic outflows are efficient in displacing gas from
galactic halos at relatively high redshift, $z\magcir 2$, when they
provide an important contribution to the enrichment of the
inter-galactic medium (IGM; e.g.,
\citealt{Oppenheimer2008MNRAS.387..577O}). At the same time, this
feedback is not efficient to quench cooling of enriched gas at low
redshift. As a consequence, no much enriched gas is left to be
stripped by the hot cluster atmosphere from the halos of merging
galaxies, thus explaining the lower enrichment level beyond
$0.1R_{180}$.

As for the runs with AGN feedback, they produce a shape on the
abundance profiles quite close to the observed ones, with a flattening
beyond $\simeq 0.2R_{180}$. In this case, the effect of AGN feedback
is that of displacing large amounts of enriched gas from star forming
regions at high redshift \citep[see
also][]{Bhattacharya2008MNRAS.389...34B} and, at the same time, to
efficiently suppress cooling at low redshift. The fact that the level
of $Z_{Fe}$ is almost constant out to $R_{180}$ and beyond, witnesses
that the main mechanism responsible for enrichment in this case is not
ram--pressure stripping, whose efficiency should decline with
cluster-centric radius. Instead, enrichment is this case is dominated
by the diffuse accretion of pre-enriched IGM. We note that the AGN2
scheme tends to predict slightly higher $Z_{Fe}$ values that
AGN1. This is due to the effect of the more efficient radio-mode
feedback, included in the former scheme, which provides a more
efficient removal of gas from the halos of massive galaxies.

Although models with AGN feedback produce the correct shape of the
Iron abundance profiles, their normalisation is generally higher than
for the observed ones. This overproduction of Iron could be due to the
uncertain knowledge of a number of ingredients entering in the
chemical evolution model implemented in the simulation code. For
instance, differences between different sets of stellar yields turn
into significant differences in the resulting enrichment level
\citep[e.g.,][]{Tornatore2007MNRAS.382.1050T,Wiersma2009arXiv0902.1535W}.
Furthermore, a reduction of the Iron abundance can also be achieved by
decreasing the fraction of binary systems, which are the progenitors
of SNe-Ia \citep[e.g.,][]{Fabjan2008MNRAS.386.1265F}. For these
reasons, we believe that the shape of the abundance profiles, instead
of their amplitude, should be considered as the relevant observational
information to be used to study the impact that different feedback
mechanisms have on the ICM enrichment pattern.

In the right panel of Fig. \ref{Fig:met_profs} we compare the average
\Zfe profiles of the five simulated galaxy groups with observational
results from the analysis of $15$ nearby galaxy groups observed with
Chandra \citep{Rasmussen2007MNRAS.380.1554R}. Also in this case, the
profiles from simulations have a slope quite similar to the observed
one out to $R\simeq 0.3R_{500}$, although with a higher normalisation.
At larger radii, the effect of AGN feedback is again that of providing
rather flat profiles. This result is at variance with the observed 
profiles in the outermost radii. Indeed, differently from
rich clusters, galaxy groups apparently show a negative gradient of
Iron abundance out to the largest radii covered by observations, with
no evidence of flattening. This result further demonstrates the
relevance of pushing observational determinations of the ICM
enrichment out to the large radii, which it is mostly sensitive to the
nature of the feedback mechanism.  If confirmed by future
observations, this result may indicate that AGN feedback needs to be
mitigated at the scale of galaxy groups, for it not to displace too
large amounts of enriched gas.

\begin{figure*}
\hbox{
\psfig{figure=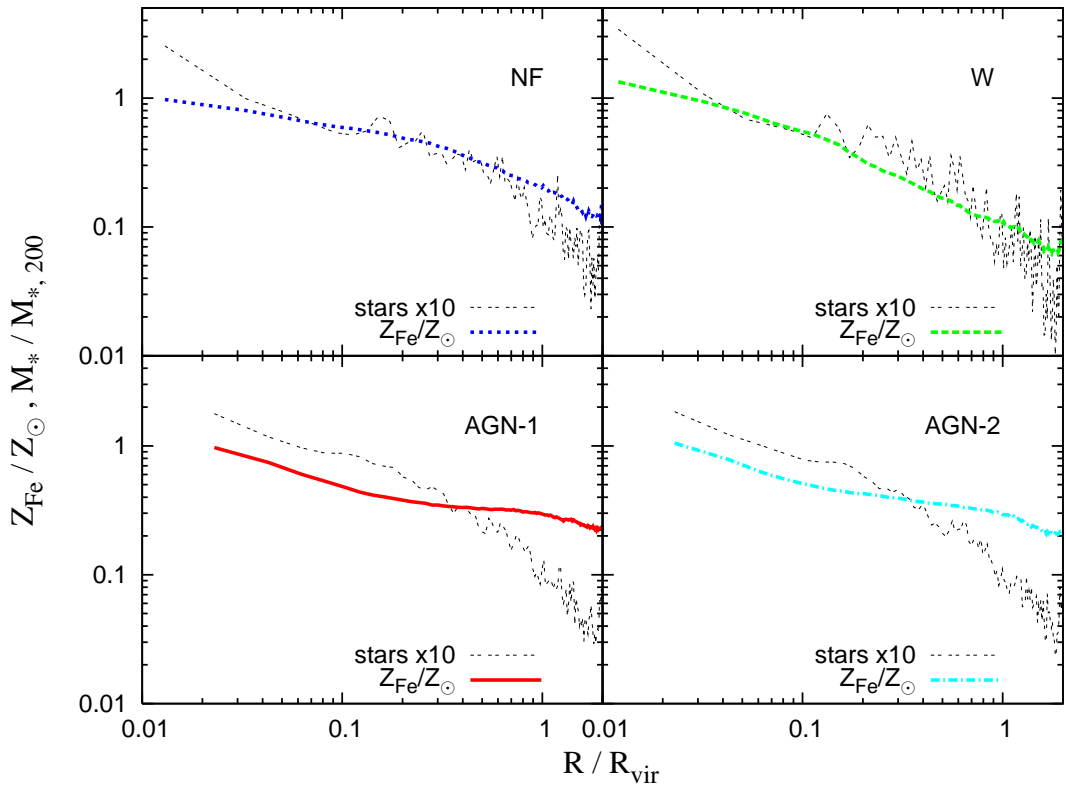,width=17.0cm,angle=-90}
}
\caption{Comparison between the stellar mass density profiles,
  normalised to the corresponding value computed within $R_{200}$
  (black short dashed lines), and the Iron abundance profiles (coloured
  lines).  For reasons of clarity stellar profiles are shifted
  by a factor $10$. All curves refer to the
  average profiles computed over the four massive clusters. From top left to
  bottom right panels we show results for the runs with no feedback
  (NF, blue short dashed), with galactic winds (W, green long dashed),
  with standard AGN feedback (AGN1, red solid) and modified AGN
  feedback (AGN2, cyan dot-dashed).}
\label{Fig:starprof}
\end{figure*}

As pointed out by \cite{Rebusco2005MNRAS.359.1041R}, the central
$Z_{Fe}$ peak in cool-core (CC) clusters (see also
\citealt{DeGrandi2004A&A...419....7D}) should be closely related to
the star (light) distribution of the BCG. On the other hand,
differences between the stellar mass profile and the metal abundance
profile should be the signature of dynamic processes which mix and
transport metals outside the BCG. In this way,
\cite{Rebusco2005MNRAS.359.1041R} derived the amount of diffusion, to
be ascribed to stochastic gas motions, which is required to explain
the shallower profiles of $Z_{Fe}$ with respect to the BCG luminosity
profiles. \cite{Roediger2007MNRAS.375...15R} carried out simulations
of isolated cluster--sized halos in which bubbles of high-entropy gas
are injected to mimic the effect of AGN feedback. They showed that the
gas diffusion associated to the buoyancy of such bubbles is indeed
able to considerably soften an initially steep metallicity profile.

In order to verify to what extent gas dynamical processes, associated
either to the hierarchical cluster build-up or to feedback energy
release, are able to diffuse metals, we compare in Figure
\ref{Fig:starprof} the average \Zfe profiles and the stellar mass
profiles for the massive clusters. As for the NF and W runs, we
clearly see a central peak in the stellar mass density profile,
related to the BCG, followed by a more gentle decline, which traces a
diffuse halo of intra-cluster stars surrounding the BCG
\citep[e.g.,][]{Murante2007MNRAS.377....2M}. Quite clearly, the \Zfe
profiles are flatter than the distribution of stars at $R\mincir 0.1
R_{vir}$. Since no feedback processes are at work in the NF runs, the
flatter \Zfe profile is due to the effect of selective cooling of
highly enriched gas, rather than to stochastic gas motions.  The same
argument can be applied also to the runs including galactic winds (W),
for which the kinetic energy provided galactic ejecta is not enough to
fully regulate star formation in central cluster regions (see also
Fig. \ref{Fig:fgas}). At intermediate radii, $R\simeq
(0.1-0.5)R_{vir}$, profiles of Iron abundance and stellar mass have
quite similar slopes in both the NF and W runs. At even larger radii,
instead, the \Zfe profile becomes again shallower than the stellar
mass profile, a trend which persists even beyond the virial radius. At
such large radii the shallower slope of \Zfe can not be explained by
the effect of gas cooling. It is rather due to the effect of gas
dynamical processes which help mixing the enriched gas. 

As for AGN feedback, its effect is instead that of providing a similar
slope for \Zfe and stellar mass profiles at $R\mincir
0.2R_{vir}$. This similarity is due to two concurrent effects. The
first one is the suppression of star formation, which allows now
central enriched gas to be pressurised by the AGN feedback, so as to
leave a relatively larger amount of metal-enriched gas in the hot
phase. The second one is the shape of the stellar mass profile, which
is less concentrated in the presence of AGN feedback. While the two
profiles are similar in the central regions, the \Zfe profile is
instead much shallower at larger radii, $R\magcir 0.3 R_{vir}$. Even
an efficient AGN feedback can not be able to displace at low redshift
significant amounts of enriched gas from the central regions of
massive clusters beyond $R_{vir}$. Therefore, the flatter abundance
profiles in the outskirts of rich clusters can only be justified by
the action played by AGN feedback at high redshift. In fact, at
$Z\magcir 2$ AGN efficiently removed enriched gas from galaxies and
quenched star formation, thus preventing metals from being locked back
in the stellar phase.

\subsection{When was the ICM enriched?}

\begin{figure}
\hbox{
\psfig{figure=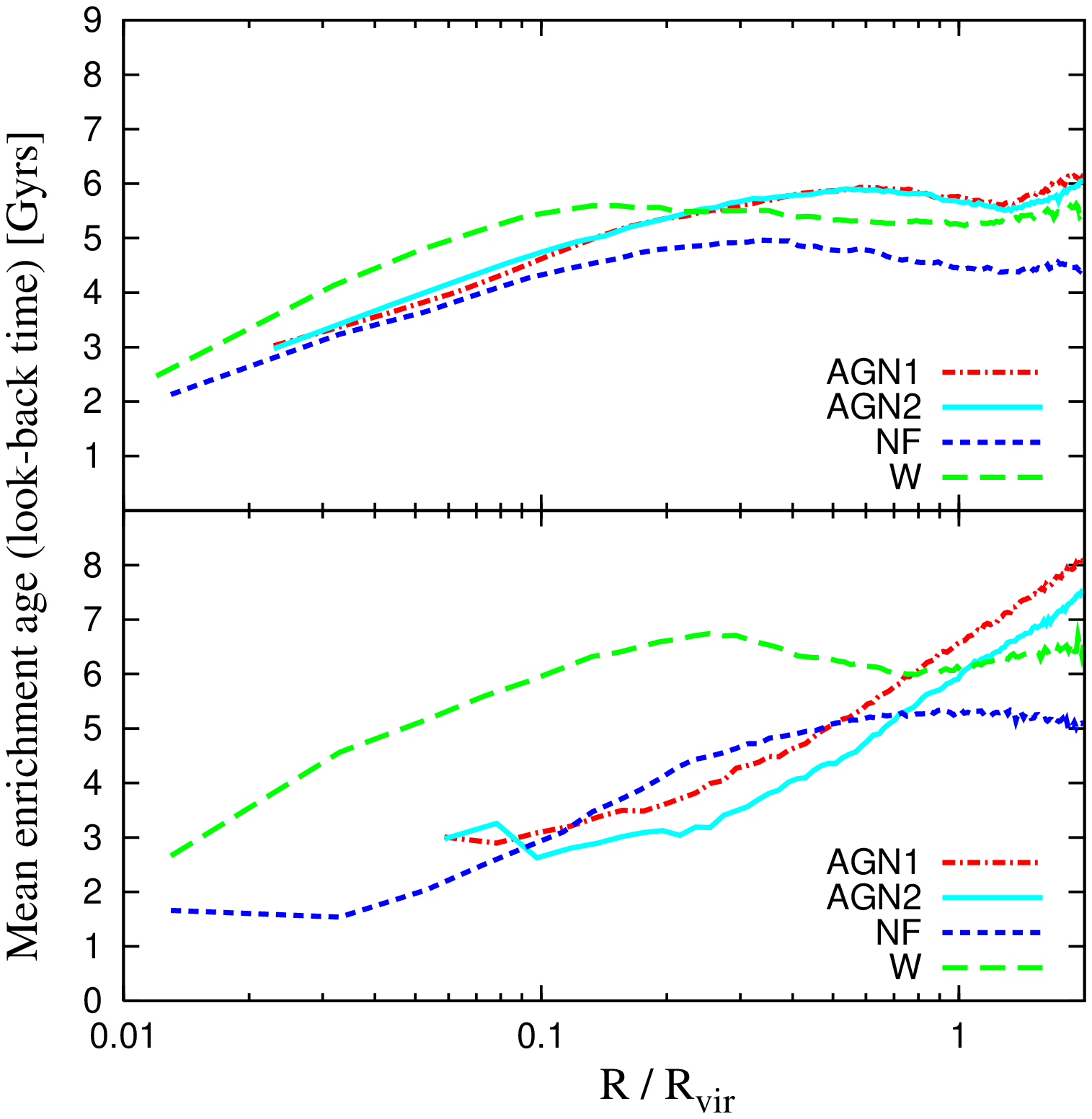,width=9.cm,angle=-90}
}
\caption{Average age of enrichment as a function of the
  cluster-centric distance. The y-axis is for the look-back time at
  which the ICM was enriched (in Gyr). The top panel is for the four
  massive clusters with $M_{200}>10^{15}\msun$, while the bottom panel
  is for the five galaxy groups with $M_{200}\simeq
  10^{14}\msun$. Different lines corresponds to the average age of
  enrichment computed within the set of simulated cluster: no feedback
  (NF, short dashed), galactic winds (W, long dashed), standard AGN
  feedback (AGN1, dot-dashed), modified AGN feedback (AGN2,
  solid). }
\label{Fig:age}
\end{figure}

In this section we discuss how the different feedback mechanisms
change the cosmic epoch at which ICM was enriched. To this
purpose, we define the average age of enrichment of a gas particle at
redshift $z$ as
\begin{equation}
\bar t(z)\,=\,{\sum_i \Delta m_{Z,i} t_i\over m_Z(z)}\,,
\label{eq:avenr}
\end{equation}
where the sum is taken over all time-steps performed by the simulation
until redshift $z$, $\Delta m_{Z,i}$ is the mass in metals received by
the particle at the $i$-th time-step, $t_i$ is the cosmic time of that
time-step and $m_Z(z)$ is the total metal mass received by the
particle before $z$. According to this definition, a large value of
enrichment age, at a given redshift, corresponds to more recent
enrichment, while smaller values of $\bar t(z)$ indicate more
pristine enrichment. In the limit in which all the metals are received
by a particle at the considered redshift $z$, then the enrichment age
coincides with the cosmic age at $z$.  Once computed for each gas
particle, we then compute the mass-weighted mean of these ages of
enrichment taken over all the particles having non--vanishing
metallicity.  

Figure \ref{Fig:age} shows how the enrichment age of the ICM changes
with the cluster-centric distance for the different feedback models,
for both rich clusters (top panel) and for poor clusters (bottom
panel).  In all cases, we note a decline towards small radii, although
with different slopes. This demonstrates that gas in central regions
has been generally enriched more recently than in the outskirts. In
the innermost regions, the typical age of enrichment correspond to a
look-back time of about 2--3 Gyr, while increasing to 5--8 Gyr around
the virial radius. This confirms that metal enrichment in the central
cluster regions receives a relatively larger contribution by star
formation taking place in the BCG, and by stripping of enriched gas
from infalling galaxies, whose star formation has been
``strangulated'' only recently by the action of the hot cluster
atmosphere. On the contrary, enrichment in the outskirts has a
relatively larger contribution from high--$z$ star formation, which
provides a more widespread IGM enrichment.
 
As for massive clusters, we note that the effect of galactic winds (W
runs) is that of providing an earlier enrichment with respect to the
model with no efficient feedback (NF runs), with a difference of about
1 Gyr, at all radii. This is quite expected, owing to the efficient
action that winds play in displacing metals from star forming regions
at high redshift. The difference between W and NF runs is larger for
the less massive clusters, consistent with the expectation that winds
are more efficient in transporting metals outside shallower potential
wells.

Quite interestingly, AGN feedback provides a steeper radial dependence
of the enrichment age, with rather similar results for the two
alternative schemes of implementation (AGN1 and AGN2). In the central
regions it is comparable to that of the NF runs, while becoming larger
than that of the W runs for $R\magcir 0.3R_{vir}$ ($\magcir
0.6R_{vir}$) for rich (poor) clusters. The relatively recent
enrichment age in central regions may look like a paradox, owing to
the effect that AGN feedback has in suppressing low redshift star
formation within the BCG (see Fig.\ref{Fig:SFR}). However, we should
remind that an aside effect of suppressing star formation is also that
of preventing recently enriched gas from being locked back in
stars. Therefore, although metal production in central regions is
suppressed in the presence of AGN feedback, this effect is compensated
by the suppression of cooling of recently enriched gas. The effect of
AGN feedback in providing a more pristine metal enrichment becomes
apparent in the outskirts of clusters. Indeed, the IGM in these
regions has been efficiently enriched at high redshift, $z\simeq
3$--4, when BH accretion reached its maximum activity level and
displaced enriched gas from star forming galaxies. After this epoch,
little enrichment took place as a consequence of the rapidly declining
star formation, thus justifying the older enrichment age. Consistently
with this picture, we also note that the increase of enrichment age
with radius is more apparent for poorer clusters, where AGN feedback
acted in a more efficient way.

\subsection{The metallicity - temperature relation}

\begin{figure*}
\hbox{
\psfig{figure=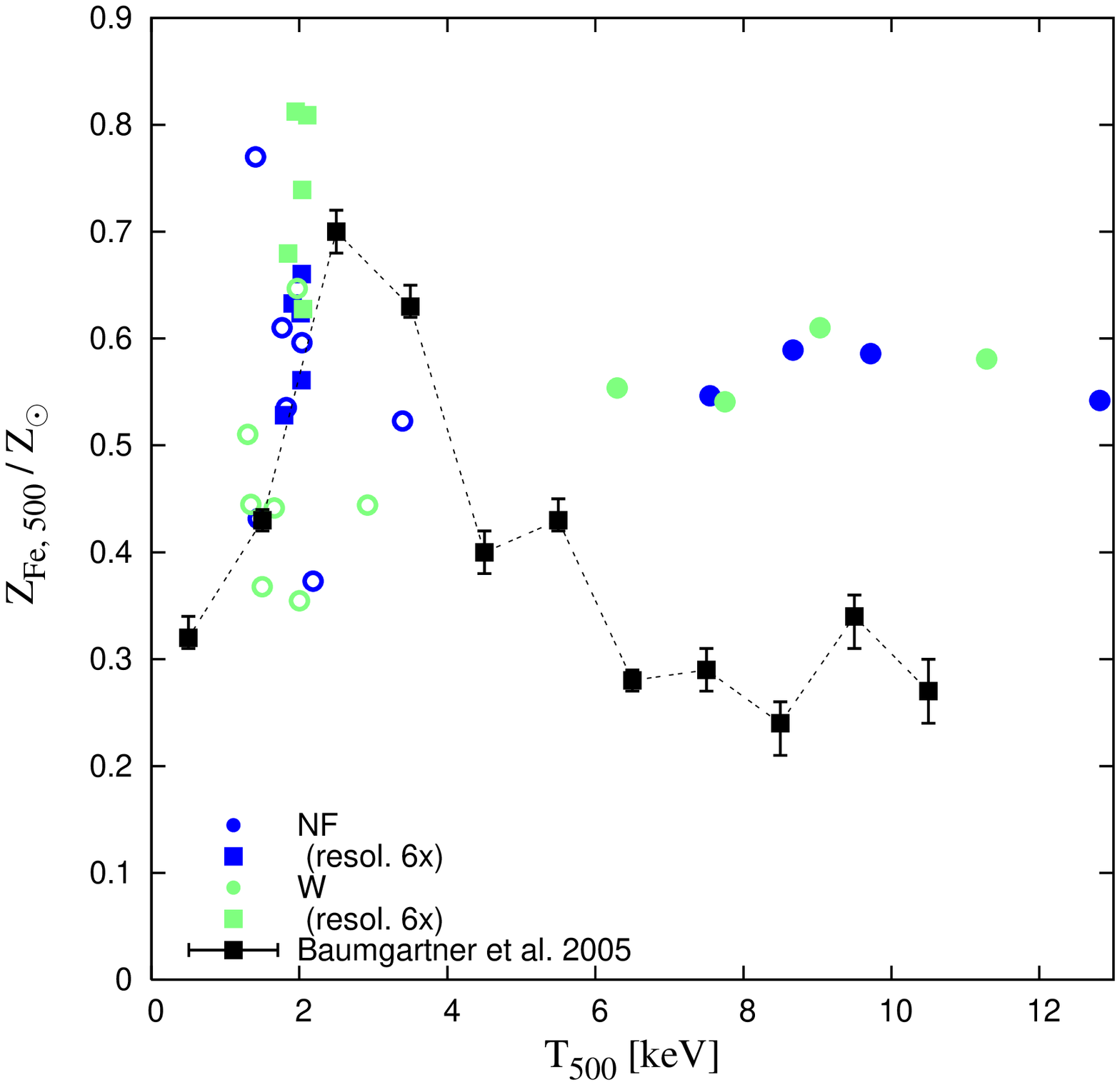,width=9.0cm,angle=-90}
\psfig{figure=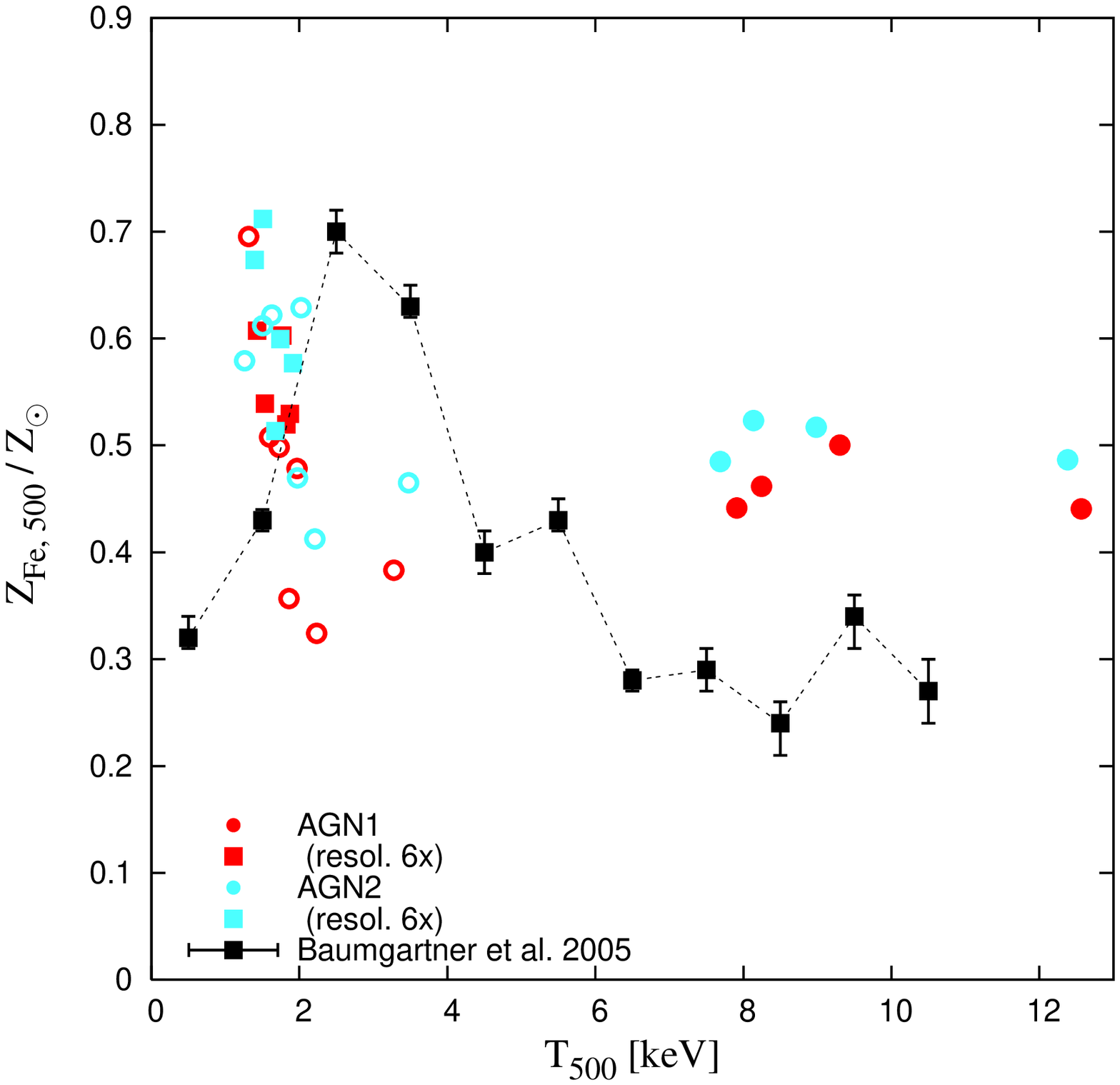,width=9.0cm,angle=-90}
}
\caption{Comparison between observed (solid squares with
  errorbars) and simulated (coloured symbols) relation between
  global Iron abundance and temperature.  The left panel shows the
  results for the NF (dark blue) and W (light green) runs, while the
  right panel shows the results for the AGN1 (dark red) and AGN2
  (light cyan) runs. In all cases, filled and open circles refer to
  the main halos of the resimulated Lagrangian regions with massive
  clusters and to the satellites, respectively. The main-halo galaxy
  groups are instead plotted with filled squares. Emission--weighted
  Iron abundance and spectroscopic--like temperature for simulations
  are both computed within $R_{500}$. Observational results refer to
  the sample of clusters observed with ASCA and analysed
  by \protect\cite{Baumgartner2005ApJ...620..680B}.}
\label{Fig:T_Z}
\end{figure*}

In Figure \ref{Fig:T_Z} we present results on the abundance of Iron as
a function of cluster temperature, by comparing simulation results for
the different feedback schemes with observational results from the
analysis of the ASCA Cluster Catalogue (ACC;
\citealt{Horner2001PhDT........88H}) carried out by
\cite{Baumgartner2005ApJ...620..680B}. Due to the relatively poor
angular resolution of the ASCA satellite, the extraction region for
each cluster was selected to contain as much flux as possible. Since a
unique extraction radius is not defined for the observed catalogue, we
adopt $R_{500}$ as a common extraction radius to compute Iron
abundances and temperatures of simulated clusters. We verified that
adopting instead a larger extraction radius (e.g. $R_{vir}$) slightly
lowers the spectroscopic-like temperatures without changing
substantially the results on \Zfe. In their analysis,
\cite{Baumgartner2005ApJ...620..680B} partitioned their large sample
of clusters in temperature bins. Then, a global value of \Zfe was
computed by combining all the clusters belonging to the same
temperature bin. In this way, errorbars associated to the
observational data points shown in Fig. \ref{Fig:T_Z} only account for
the statistical uncertainties in the spectral fitting procedure after
combining all clusters belonging to the same temperature bin, while
they do not include any intrinsic scatter (i.e., cluster-by-cluster
variation) in the \Zfe--$T$ relation. This has to be taken in mind
when comparing them to simulation results, for which we did not make
any binning in temperature.

In both panels we see that simulations of hot ($T> 6$ keV) systems
produce values of \Zfe which are above the enrichment level found in
observations, $\sim 0.3 Z_{\sun}$. This result is consistent with the
fact that \Zfe profiles for simulated clusters have a larger
normalisation than the observed ones (see
Fig. \ref{Fig:met_profs}). NF and W runs predict $Z_{Fe}\simeq
(0.5-0.6) Z_{Fe,\sun}$, with a slightly lower value, $Z_{Fe}\simeq
(0.4-0.5) Z_{Fe,\sun}$ for the runs with AGN feedback.

As for the simulations of the poorer systems, with $T\simeq (1-3)$
keV, they have instead a larger scatter. This indicates more diversity
in the impact that gas-dynamical and feedback processes have in
determining the enrichment pattern within smaller systems.  The effect
of AGN feedback for these systems is that of decreasing the value of
\Zfe by about $0.1Z_{Fe,\odot}$. The rather large spread of abundance
values and the limited number of simulated systems do not allow to
establish whether simulations reproduce the increase of \Zfe with ICM
temperature for systems in the range $T\simeq (1-3)$ keV.

Observations suggests that \Zfe is almost independent of temperature
above 6 keV, a trend which is in fact reproduced by simulation
results. However, between 3 and 6 keV observed clusters show a drop in
metallicity by about a factor two. Although simulations predict
somewhat higher metallicity values at low temperatures, still it is
not clear whether they reproduce the decrease, by more than a factor
2, suggested by observations. A potential complications in comparing
observational and simulation results for systems with $T\sim 3$ keV is
that the spectroscopic value of \Zfe for these systems is contributed
by both K and L lines. As originally noted by
\cite{Buote2000MNRAS.311..176B} in the analysis of ASCA data, fitting
with a single-temperature model a plasma characterised by a
multi-temperature structure, with the colder component below 1 keV,
leads to an underestimate of the Iron abundance \citep[the so-called
iron-bias; see
also][]{Molendi2001A&A...375L..14M,Buote2003ApJ...595..151B}. In this
case, a simple emission--weighted definition of \Zfe from simulations
may not be fully adequate. A correct procedure would require
extracting a mock X--ray spectrum from simulated clusters, to be
fitted with a multi--temperature (and multi--metallicity) plasma model
\citep[e.g.,][]{Rasia2008ApJ...674..728R}. Finally, one should also
note that more recent determinations of the \Zfe--$T$ relation from
XMM--Newton data, although based on a much smaller number of clusters,
suggests a less pronounced decrease for systems hotter than 3 keV
\citep[e.g.,][]{Werner2008SSRv..134..337W}. There is no doubt that a
systematic analysis of nearby clusters within the Chandra and
XMM-Newton archives would help to confirm or disprove the
metallicity--temperature relation based on ASCA observations.

\subsection{The \Zsife relative abundance}
The relative abundance of elements produced in different proportions
by different SN types is directly related to the shape and possible
evolution of the initial mass function. Furthermore, studying how
relative abundances change with cluster-centric distance provides
insights on the different timing of enrichment and on how different
metals, produced over different time-scales, are mixed by
gas-dynamical processes. ASCA data analysed by
\cite{Loewenstein1996ApJ...466..695L},
\cite{Fukazawa1998PASJ...50..187F} and
\cite{Finoguenov2000ApJ...544..188F} originally suggested that cluster
outskirts are predominantly enriched by SNe-II. A similar result was
found more recently also by \cite{Rasmussen2007MNRAS.380.1554R}, who
analysed XMM data for poor clusters with $T\mincir 3$ keV. On the
other hand, Suzaku observations of low temperature clusters and groups
(\citealt{Sato2008PASJ...60S.333S}, \citeyear{2009PASJ...61S.353S}a,
\citeyear{Sato2009PASJ...61S.365S}b) show instead a rather flat
profile of $Z_{Si}/Z_{Fe}$ out to large radii, $\simeq 0.3 R_{vir}$,
thus implying that SNe-Ia and SNe-II should contribute in similar
proportions to the enrichment at different radii.

\begin{figure*}
\hbox{
\hspace{1.5truecm}
\psfig{figure=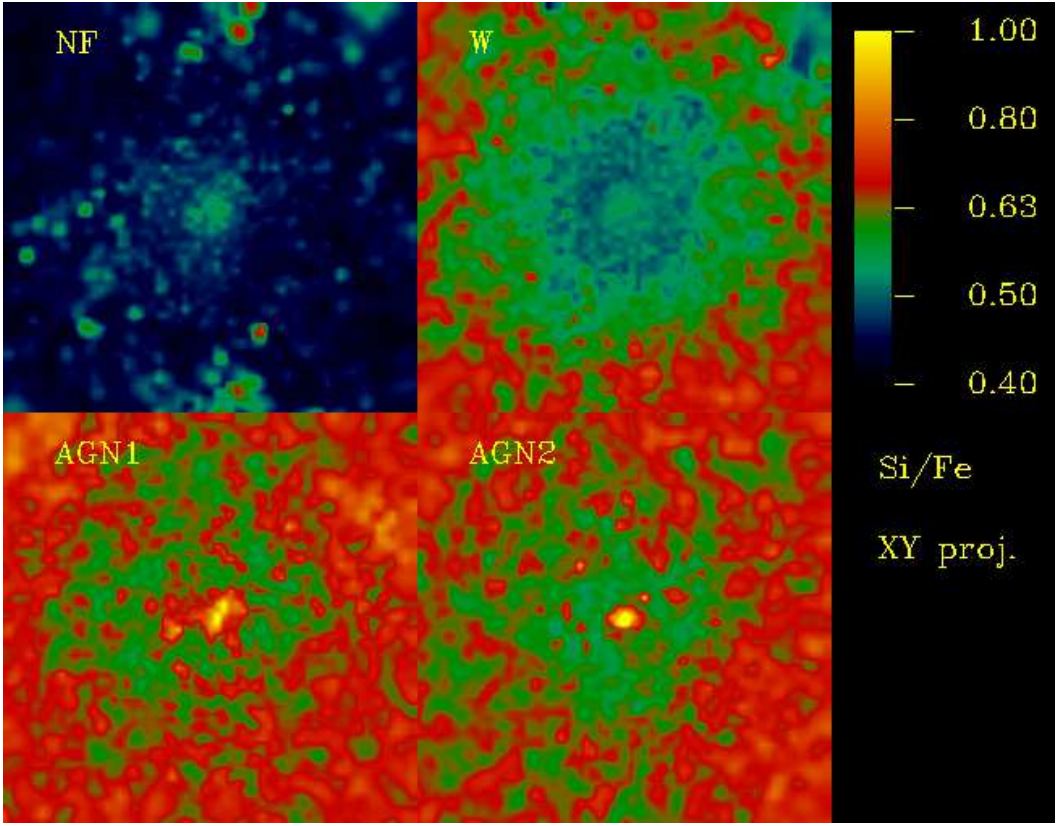,width=14.0cm,angle=0}
}
\caption{Maps of the emission--weighted \Zsife distribution for the
  runs of the g676 cluster without feedback (NF, top left), with winds
  (W, top right), with AGNs (AGN1 and AGN2, bottom left and right,
  respectively).  The side of each map is $2 R_{vir}$. Abundance
  values are expressed in units of the solar value, as reported by
  \protect\cite{Grevesse1998SSRv...85..161G}, with color coding
  specified in the right bar.}
\label{Fig:maps2}
\end{figure*}

In this section, we focus our attention on the \Zsife ratio for low
temperature clusters. We show in Figure \ref{Fig:maps2} the
emission-weighted map of the \Zsife ratio for the four different
feedback schemes applied to the g676 cluster. For the run with no
feedback (NF), we note that \Zsife is generally quite patchy. It
reaches higher values in correspondence of high--density star forming
regions, a feature which is also shared in different proportions by
the other runs. Indeed, the products of SNe-II are released over a
short time scale, since they are synthesised by massive stars. As a
result, their distribution tends to trace preferentially the
distribution of star-forming regions. On the other hand, SNe-Ia
release metals over a longer time-scale. In fact, these stars have
time to leave star forming regions, as a consequence of the same
dynamical effects which generate a population of inter-galactic stars
\citep[e.g.,][]{Murante2007MNRAS.377....2M,Zibetti2005MNRAS.358..949Z},
thereby providing a more widespread enrichment pattern. Therefore, our
simulations predict that diffuse intra-cluster stars provide a
significant contribution to the enrichment of the intra-cluster medium
\citep[see also][]{Tornatore2007MNRAS.382.1050T}.

A comparison of the maps for the NF and W runs shows that they have
comparable levels of \Zsife within the central regions. In these
regions we expect that ram--pressure stripping is the dominant process
in removing gas from merging galaxies \citep[see also
][]{Kapferer2007A&A...466..813K} and, therefore, in efficiently mixing
in the ICM the nucleosynthetic products of different stellar
populations. On the contrary, in the cluster outskirts winds have been
much more efficient in removing freshly produced metals from galaxies
during the cluster assembly, therefore providing a relatively more
widespread Si distribution, with respect to the NF case in which no
galactic outflows are included.
 
The runs with AGN feedback exhibit a behaviour in the cluster
outskirts which is qualitatively similar to that of the W run,
although with slightly higher values of \Zsife. This indicates that
AGN feedback has a higher efficiency in mixing SN-Ia and SN-II
products. As for the central regions, the truncation of recent star
formation by AGN feedback would lead to the naive
expectation that a relatively higher value of Iron abundance with
respect to Silicon should be found. The results shown
in the bottom panels of Fig.\ref{Fig:maps2} lead in fact to the
opposite conclusion, with a marked increase of \Zsife in the core
regions. The reason for this lies again in the effect of selective
removal of metal-enriched gas associated to cooling. Total metallicity
of the gas around the BCG is dominated by SNe-II
products. Therefore, gas more enriched by SNe-II has a
relatively shorter cooling time. As a consequence, suppression of
cooling in the core regions by AGN feedback tends to increase the
amount of SN-II products in the ICM, thereby justifying the increase
of \Zsife with respect to the runs not including AGN.

\begin{figure*}
\hbox{
\psfig{figure=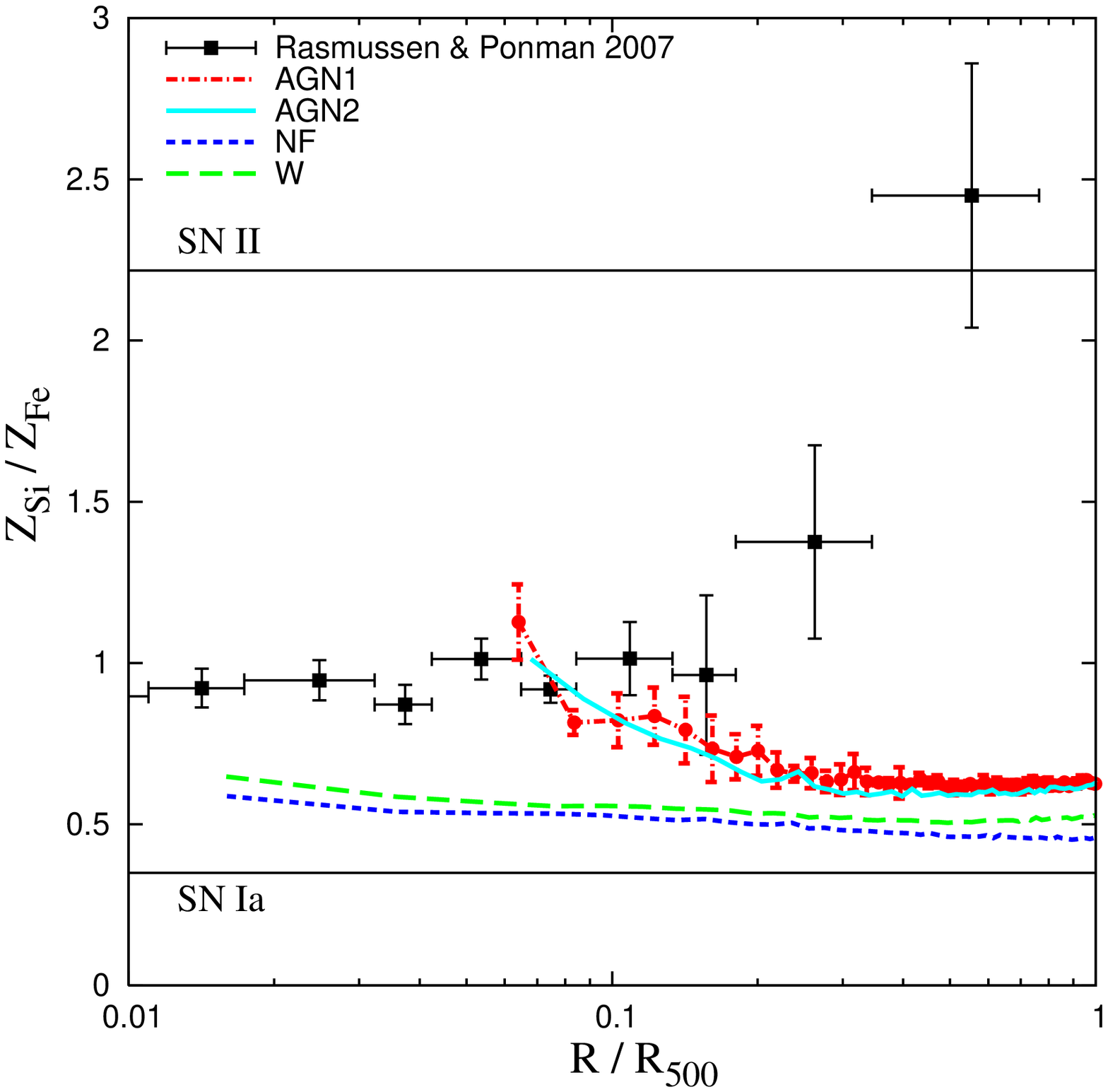,width=9.0cm,angle=-90}
\psfig{figure=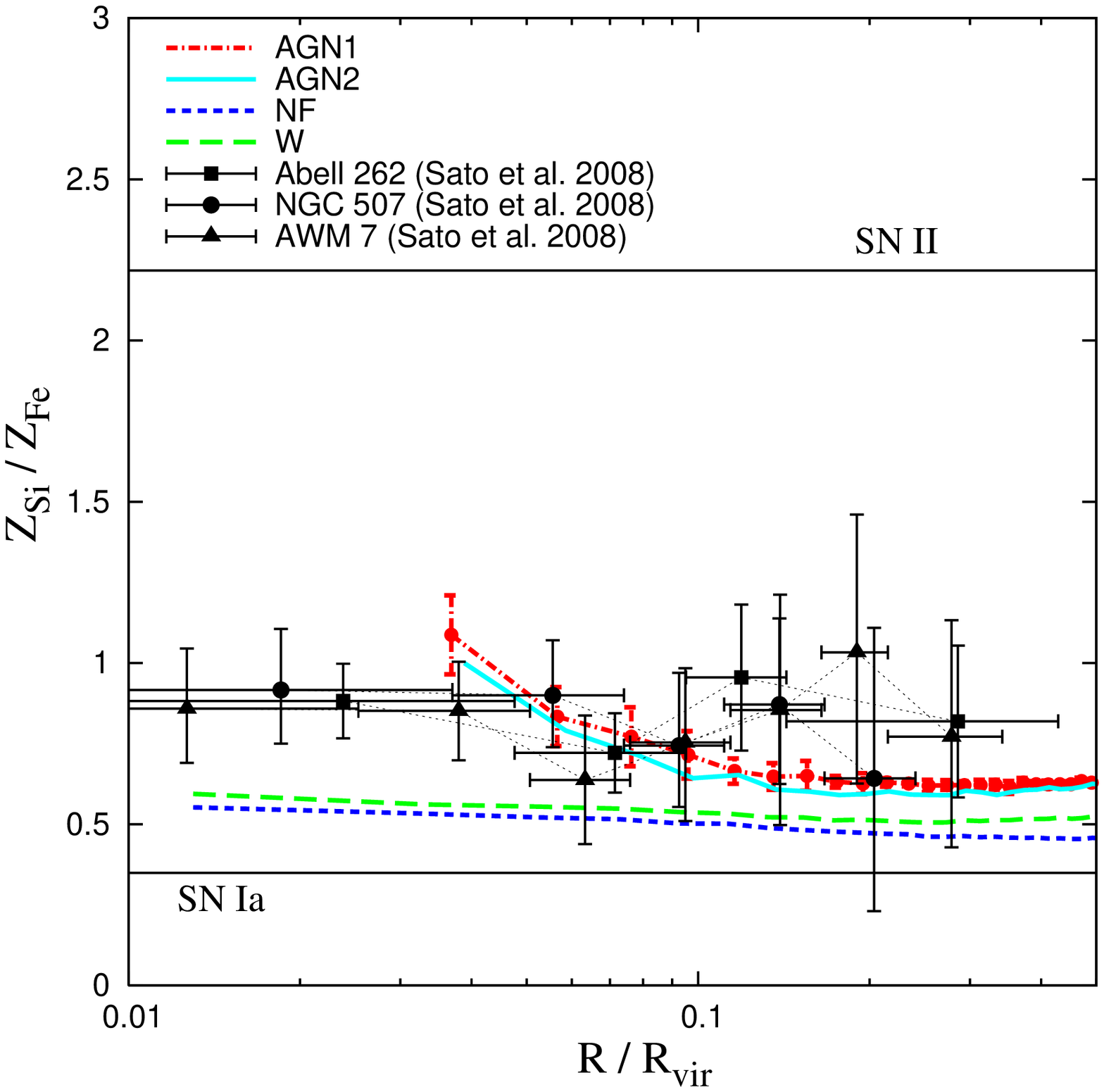,width=9.0cm,angle=-90}
}
\caption{Comparison between observed and the simulated profiles of
  Silicon abundance relative to Iron, \Zsife. In each panel, different
  lines correspond to the average profiles computed over clusters
  having $T_{500}<3$ keV for the different sets of runs: no feedback
  (NF, blue short dashed), galactic winds (W, green long dashed),
  standard AGN feedback (AGN1, red dot-dashed), modified AGN feedback
  (AGN2, cyan solid). For reasons of clarity, we show with errorbars
  the r.m.s. scatter over the ensemble of simulated clusters only for
  the AGN1 runs. Observational points refer to the Chandra data
  analysed by \protect\cite{Rasmussen2007MNRAS.380.1554R} (left panel)
  and to the Suzaku data analysed by
  \protect\citeauthor{Sato2008PASJ...60S.333S} 
  (\protect\citeyear{Sato2008PASJ...60S.333S},
   \protect\citeyear{2009PASJ...61S.353S}a,
   \protect\citeyear{Sato2009PASJ...61S.365S}b)
   (right panel). The two horizontal lines
  show the relative abundance from SNe-Ia and SNe-II, computed for a
  simple stellar population (SSP) having solar initial metallicity and
  based on the same Salpeter IMF and set of yields as used in our
  simulations.}
\label{Fig:SiFe_profs}
\end{figure*}

This qualitative picture is also confirmed by the profiles of \Zsife,
that are plotted in Figure \ref{Fig:SiFe_profs}. In each panel,
simulation results show the average profile, computed over the 5 main
relatively poor clusters, for each feedback model. In the left panel
simulation results are compared with observational results by
\cite{Rasmussen2007MNRAS.380.1554R} from the analysis of $15$ nearby
galaxy groups observed with Chandra. Also shown with the two
horizontal lines are the values of \Zsife produced by SNe-Ia and SNe-II
for a simple stellar population of initial solar metallicity, having a
Salpeter IMF, using the same sets of yields adopted in our simulations.

Observational data show a rather flat profile of \Zsife with a value
close to solar at small radii, followed by a sudden increase beyond
$\simeq 0.2 R_{500}$. Taken at face value, this result would imply
that at $\sim 0.5R_{500}$ the enrichment is mostly contributed by
SNe-II, while a mix of different stellar populations is required in the
cluster centre. Although this result is qualitatively similar to that
suggested by the visual inspection of the \Zsife maps of
Fig.\ref{Fig:maps2}, simulation results are quantitatively different
from those by \cite{Rasmussen2007MNRAS.380.1554R}. Indeed, in no case
simulations predict an increase of $Z_{Si}$/$Z_{Fe}$ beyond $0.2
R_{500}$. While the maps suggest that such an increase is also
expected in simulations, it takes place only for $R>R_{500}$, with the
largest values reaching at most the solar one. Within the radial range
covered by Chandra observations, both NF and W runs show rather flat
profiles. Also in this case, we are more interested in the slope of
the \Zsife profile, rather than in its normalisation. Indeed, too low
values of $Z_{Si}$/$Z_{Fe}$ in simulations by about 0.3 (in solar
units) can be be either due to the choice of the IMF or to adopted
stellar yields.

As for the runs with AGN feedback the profiles of $Z_{Si}$/$Z_{Fe}$
confirm the expectation gained from the maps of Fig.\ref{Fig:maps2} for a
relative increase of the Si abundance in the central regions. As for
the behaviour at large radii, $R\magcir 0.2R_{500}$, also AGN feedback
does not predict the pronounced increase see in the Chandra data by
\cite{Rasmussen2007MNRAS.380.1554R}.

As already mentioned, this observational evidence for an enhancement
of SN-II products in the outskirts of groups is not confirmed by
Suzaku data (\citealt{Sato2008PASJ...60S.333S},
\citeyear{2009PASJ...61S.353S}a,\citeyear{Sato2009PASJ...61S.365S}b)
and XMM-Newton (Silvano Molendi, private communication). As shown in
the left panel of Fig.\ref{Fig:maps2}, Suzaku observations of Abell
262, NGC 507 and AWM 7 indicate a profile of \Zsife which is
consistent with being flat out to $0.3 R_{vir}$ (corresponding to
about $0.5R_{500}$). While it is not the purpose of this work to
address the reason for the different results coming from different
satellites, we want to stress the great relevance of tracing the
pattern of ICM enrichment out to the largest possible radii. Indeed,
this is the regime where gas-dynamical processes related to the
cosmological build-up of clusters, past history of star formation and
nature of feedback processes regulating star formation, all play a
role in determining the distribution of metals.

\vspace{0.2truecm} In summary, the analysis of chemical enrichment in
our simulations of galaxy clusters confirms that the resulting
metallicity distribution in the ICM is given by the interplay between
gas cooling, which tends to preferentially remove more enriched gas
from the hot phase, feedback processes, which displace gas from star
forming regions and regulate star formation, and gas dynamical
processes associated to the hierarchical build-up of galaxy
clusters. In particular, feedback implemented through the action of
galactic winds powered by SN explosions or through energy extracted
from gas accretion onto super-massive BHs, leave distinct features on
the resulting pattern and timing of ICM enrichment.

\section{Conclusions}
We presented the analysis of an extended set of cosmological
hydrodynamical simulations of galaxy clusters aimed at studying the
different effects that stellar and AGN feedback have on the thermal and
chemo-dynamical properties of the intra-cluster medium (ICM). Using a
version of the Tree-SPH \gadget code
\citep{Springel2005MNRAS.364.1105S}, which also includes a detailed
description of chemical enrichment
\citep{Tornatore2007MNRAS.382.1050T}, we carried out simulations of 16
clusters identified within 9 Lagrangian regions extracted from a
lower-resolution parent cosmological box
\citep{Dolag2008arXiv0808.3401D}. All cluster simulations of this set
have been run using different prescriptions for the feedback: without
including any efficient feedback (NF runs), including only the effect
of galactic winds powered by supernova (SN) feedback (W runs), and
including two different prescriptions of AGN feedback (AGN1 and AGN2)
based on modelling gas accretion on super-massive black holes (BHs) hosted
within resolved galaxy halos
\citep{Springel2005MNRAS.361..776S,DiMatteo2005Natur.433..604D}. The
AGN1 scheme exactly reproduces the original model by
\citep{Springel2005MNRAS.361..776S} for the choice of the parameters
determining the feedback efficiency and the way in which energy is
distributed. As for the AGN2 scheme, it assumes the presence of a
radiatively efficient ``radio mode'' phase when BH accretion is in a
quiescent stage \citep[e.g.,][]{Sijacki2007MNRAS.380..877S}, also
distributing energy to the gas particles surrounding BHs in a more
uniform way.

The main results of our analysis can be summarised as follows.
\begin{itemize}
\item[(a)] AGN feedback significantly quenches star formation rate
  (SFR) associated to the brightest cluster galaxies (BCGs) at
  $z\mincir 4$. At $z=0$ the SFR in the AGN1 and AGN2 models is
  reduced by about a factor six. For a massive cluster with
  $M_{200}\simeq 10^{15}\msun$ we find $SFR(z=0)\simeq
  70M_\odot$yr$^{-1}$, thus not far from current observational
  estimates \citep[e.g.][]{Rafferty2006ApJ...652..216R}. Although the
  two variants of AGN feedback produce similar results on the star
  formation rate, the AGN2 model is more efficient is reducing gas
  accretion onto BHs. For this scheme, the resulting masses of the BHs
  sitting at the centre of the BCGs at $z=0$ are reduced by a factor
  3--5 with respect to the AGN1 scheme. Furthermore, increasing
    the radio-mode feedback efficiency from $\epsilon_f=0.2$ to 0.8
    further reduces the mass of the central BH by about a
    factor 2.5, while leaving the level of low-$z$ star formation rate
    almost unaffected.
\item[(b)] AGN feedback brings the $L_X$--$T$ relation in closer
  agreement with observational results at the scale of poor clusters
  and groups, thus confirming results from previous simulations based
  on AGN feedback \citep{Puchwein2008ApJ...687L..53P}. However, this
  is obtained at the expense of increasing the ICM entropy in central
  regions of groups above the level indicated by observational results
  \citep[e.g.,][]{Sun2009ApJ...693.1142S,Sanderson2009MNRAS.395..764S}. This
  entropy excess generated in central group regions corresponds in
  turn to a too low value of the gas fraction.
\item[(c)] AGN feedback reduces by 30--50 per cent the fraction of
  baryons converted into stars, $f_{star}$, within
  $R_{500}$. Simulation results agree well with the observed value of
  $f_{star}$ at the scale of groups. However, for rich clusters the
  fraction of stars within $R_{500}$ from simulation ($\simeq 30$--40
  per cent) is larger than the observed one ($\simeq 10$ per cent).
\item[(d)] AGN feedback is quite efficient in pressurising gas in the
  central regions of galaxy groups, thereby generating temperature
  profiles which are in reasonable agreement with the observed
  ones. Despite this success at the scale of groups,
  temperature profiles in the core regions of massive clusters are
  still too steep, even after including AGN feedback.
\item[(e)] The presence of AGN feedback generates a rather uniform and
  widespread pattern of metal enrichment in the outskirts of
  clusters. This is the consequence of the improved efficiency, with
  respect to the runs without BH feedback, to extract at high redshift
  highly enriched gas from star forming regions, and, therefore, to
  enhance metal circulation in the inter-galactic medium.
\item[(f)] Radial profiles of Fe abundance are predicted to be too
  steep at $R\magcir 0.1 R_{180}$ in runs including stellar feedback.
  Their shape is in much better agreement with the observed ones when
  including AGN feedback. The overall emission--weighted level of
  enrichment within massive clusters is $Z_{Fe}\simeq 0.5$ and 0.6 for
  runs with and without AGN feedback, respectively. Such values are
  generally larger than those, $Z_{Fe}\simeq 0.3$, reported from ASCA
  observations \citep{Baumgartner2005ApJ...620..680B}.
\item[(g)] The distribution of elements mostly produced by SNe-II over
  a relatively short time-scale is more clumpy than the Iron
  distribution, which has a larger contribution from SNe-Ia, that
  release metals over a long time-scale. This is interpreted as due to
  the effect of enrichment from stars belonging to a diffuse
  intra-cluster population. Therefore, simulations predict that a
  sizable fraction of the ICM enrichment is produced by intra-cluster
  stars, in line with observational evidences
  \citep{Sivanandam2009ApJ...691.1787S}.
\item[(h)] The runs with no feedback (NF) and with galactic winds (W)
  predict similar patterns of \Zsife within $R_{500}$. Silicon
  abundance is enhanced in the outer regions by the action of galactic
  winds, thanks to their efficiency in transporting gas enriched by
  $\alpha$ elements from star forming regions. AGN feedback has the
  effect of increasing \Zsife as a consequence of the suppression of
  star formation, which would lock back in stars gas surrounding star
  forming regions, and of the efficient removal of enriched gas from
  galactic halos.
\item[(i)] In no case we find that profiles of \Zsife have a rising
  trend beyond $\simeq 0.2R_{500}$.  No strong conclusion can be drawn
  from a comparison with data, owing to discrepant indications from
  different observational results on the radial dependence of
  \Zsife. Suppression of star formation with AGN feedback causes
  \Zsife to increase at small radii, $\mincir 0.1R_{500}$, a feature
  which is not seen in observational data. This suggests that the
  implementation of AGN feedback in our simulations may not provide
  enough gas mixing the central regions of clusters and groups.
\end{itemize}

Our analysis lend further support to the idea that a feedback source
associated to gas accretion onto super-massive BHs is required by the
observational properties of the ICM
\citep[e.g.][]{McNamara2007ARA&A..45..117M}. However, our results also
show that a number of discrepancies between observations and
predictions of simulations still exist, especially within the core
regions of massive clusters. This requires that a more efficient way
of extracting and/or thermalising energy released by AGN should be
introduced in richer systems. A number of observational evidences
exists that AGN should represent the engine which regulates the
structure of core regions of clusters and groups. However,
observations also provide circumstantial evidences that a number of
complex physical processes, such as injection of relativistic
particles and of turbulence associated to jets, buoyancy of bubbles
stabilised by magnetic fields, viscous dissipation of their mechanical
energy, thermal conduction, should all cooperate to make AGN feedback
a self-regulated process. In view of this complexity, we consider it
as quite encouraging that the relatively simple prescriptions for
energy thermalisation adopted in our simulations provide a significant
improvement in reconciling numerical and observational results on the
ICM thermo- and chemo-dynamical properties.

  Clearly, increasing numerical resolution thanks to the ever
  increasing supercomputing power would require including a proper
  description of the above processes. For instance, the current
  implementation of AGN feedback neglects the
  effect of kinetic energy associated to jets. The typical scales of
  $\sim 20$ kpc at which kinetic feedback is expected to dominate
  within clusters \citep[e.g.][]{Pope2009MNRAS.395.2317P} are only
  marginally resolved by hydrodynamics in our simulations, thus making
  the assumption of a purely thermal feedback a reasonable
  one. However, the improved numerical resolution expected to be
  reached in simulations of the next generation needs to be
  accompanied by a suitable description of injection of jets, for them
  to provide a physically meaningful description of the interplay
  between BH accretion and ICM properties.

The results presented in this paper further demonstrate that different
astrophysical feedback sources leave distinct signatures on the
pattern of chemical enrichment of the ICM. These differences are much
more evident in the outskirts of galaxy clusters, which retain memory
of the past efficiency that energy feedback had in displacing enriched
gas from star-forming regions and in regulating star formation
itself. However, characterisation of thermal and chemical properties
in cluster external regions requires X--ray telescopes with large
collecting area and an excellent control of the background. While
Chandra, XMM and Suzaku will be pushed to their limits in these
studies in the next few years, there is no doubt that a detailed
knowledge of the ICM out the cluster virial boundaries has to await
for the advent of the next generation of X--ray telescopes
\citep[e.g.,][]{Giacconi2009astro2010S..90G,Arnaud2009astro2010S...4A}. 

\section*{Acknowledgements}
We are greatly indebted to Volker Springel for providing us with the
non--public version of \gadget.  We thank Anthony Gonzalez, Alberto
Leccardi, Jesper Rasmussen and Ming Sun for having kindly provided
tables of observational data points, and Silvia Ameglio for her help
in producing the metallicity maps. We acknowledge useful discussions
with Hans B\"ohringer, Francesca Matteucci, Pasquale Mazzotta, Silvano
Molendi, Ewald Puchwein, Elena Rasia, Debora Sijacki, Paolo Tozzi,
Matteo Viel and Mark Voit.  The simulations have been carried
out at the ``Centro Interuniversitario del Nord-Est per il Calcolo
Elettronico'' (CINECA, Bologna), with CPU time assigned thanks to an
INAF--CINECA grant and to an agreement between CINECA and the
University of Trieste, and at the Computing Centre of the University
of Trieste. This work has been partially supported by the INFN PD-51
grant, by the INAF-PRIN06 Grant, by a ASI-AAE Theory grant and by the
PRIN-MIUR 2007 grant. KD acknowledges the financial support by the
``HPC-Europa Transnational Access program'' and the hospitality of
CINECA.

\bibliographystyle{mn2e}
\bibliography{master}

\end{document}